\documentclass[epj]{svjour}

\usepackage{epsfig}
\usepackage{amsmath,amssymb,amsfonts}
\usepackage{latexsym}
\usepackage{graphicx}
\newcommand{\eq}{\begin{eqnarray}} 
\newcommand{\en}{\end{eqnarray}}
\newcommand{\ra}{\rangle}
\newcommand{\la}{\langle}

\begin{document}

\title{$f_0(980)$ meson as a $K\bar K$ molecule \\
in a phenomenological Lagrangian approach}
\author{Tanja Branz \and Thomas Gutsche \and Valery Lyubovitskij
\thanks{\emph{On leave of absence from the
Department of Physics, Tomsk State University,
634050 Tomsk, Russia}}%
}

\institute{Institut f\"ur Theoretische Physik,
Universit\"at T\"ubingen,
\\ Auf der Morgenstelle 14, D-72076 T\"ubingen, Germany}
\date{Received: date / Revised version: date}
%
\abstract{
We discuss a possible interpretation of the $f_0(980)$ meson 
as a hadronic molecule - a bound state of $K$ and $\bar K$ 
mesons. Using a phenomenological Lagrangian approach we calculate 
the strong $f_0(980) \to \pi\pi$ and electromagnetic 
$f_0(980) \to \gamma\gamma$ decays. The compositeness condition 
provides a self-consistent method to determine the coupling 
constant between $f_0$ and its constituents, $K$ and $\bar K$. 
Form factors governing the decays of the $f_0(980)$ are calculated by 
evaluating the kaon loop integrals. 
The predicted $f_0(980) \to \pi\pi$ and $f_0(980) \to \gamma\gamma$
decay widths are in good agreement with available data and results 
of other theoretical approaches. 
} 
\maketitle

\section{Introduction}

The understanding of the structure of scalar mesons with masses 
around 1 GeV is one of the prominent topics in modern hadronic 
physics. The study of scalar mesons can for example shed light on 
the problem of the QCD vacuum, {\it e.g.} to understand the role of gluon 
configurations and strangeness in the formation of their spectrum. 
Different interpretations of scalar mesons have been suggested and 
developed during the last decades~\cite{Yao:2006px}. 
The canonical picture is based on the constituent $q \bar q$ structure 
of scalar mesons. In this vein, by analogy with pseudoscalar and 
vector mesons, one can organize the low-lying scalar mesons, a triplet of 
$a_0(980)$, two doublets of $K_0^\ast(1430)$ and two 
singlets $f_0(980)$ and $f_0(1370)$, in the form 
of the $J^P=0^{+}$ nonet (see {\it e.g.} discussion in ref.~\cite{Palano:1994ir}). 

In this paper we focus on the $f_0(980)$ meson. Analyses of the $f_0(980)$ 
meson as a quarkonium state were performed in several papers (see, {\it e.g.} 
refs.~\cite{Efimov:1993ei}-\cite{Anisovich:2005kt}). Different scenarios 
for the admixture of nonstrange  and strange $q\bar q$ components have 
been developed, which range from a pure or dominant $s \bar s$ 
state~\cite{Efimov:1993ei}-\cite{vanBeveren:1998qe} to a dominant 
$n \bar n = (u \bar u + d \bar d)/\sqrt{2}$ 
configuration~\cite{Celenza:2000uk} with a small $s \bar s$ mixture of 
about 10\%. Extensions of this scheme by mixing of quarkonia and glueball 
components have been analyzed in refs.~\cite{Volkov:2000vy,Anisovich:2001zr}. 
In~\cite{Volkov:2000vy} it was found that the strong $f_0(980) \to \pi\pi$ 
decay width is determined by the quarkonium part, while the glueball 
contribution is small. The dominance of the quarkonium part was also 
confirmed in ref.~\cite{Anisovich:2001zr} in the analysis of the radiative 
decays of $f_0(980)$. In ref.~\cite{Jaffe:1976ig} the existence of 
scalar multiquark-states was suggested. Scalar mesons (including the 
$f_0(980)$ meson) have been assigned to the lightest cryptoexotic 
$q^2 \bar q^2$ nonet. A further development of the four-quark model for 
the $f_0(980)$ has been done in~\cite{Achasov:1981kh} and recently 
in~\cite{Giacosa:2006rg}. Properties of the $f_0(980)$ resulting from 
the $q \bar q$ and $q^2 \bar q^2$ schemes have been critically analyzed. 
In~\cite{Black:2007bp} the structure of the light scalar nonet including 
$f_0(980)$ was tested using radiative $\phi$ decays. The authors of 
Ref.~\cite{Black:2007bp} point out the difficulty to distinguish 
between the $q\bar q$ and the $q q \bar q \bar q$ picture for the light 
scalar mesons. A possible admixture between $\bar q q$ and 
$q q \bar q \bar q$ configurations for the low-lying scalar mesons
has been considered in ref.~\cite{Fariborz:2006ff} using the chiral approach. 
In refs.~\cite{Weinstein:1982gc} the idea of multi-quark states has been 
put forward to allow for the arrangement of the two quarks and two 
antiquarks as a bound state of a kaon and a antikaon. Different 
approaches describing the $f_0(980)$ as a hadronic molecule have 
already been discussed~\cite{Weinstein:1982gc}-\cite{Lemmer:2007qp}. 
The treatment of the bound state $K\bar K$ interaction ranges from simple 
Gaussian ~\cite{Barnes:1985cy} and 
meson-exchange~\cite{Krehl:1996rk,Janssen:1994wn} potentials to chiral 
perturbation theory (ChPT)~\cite{Oller:1998zr}. Besides this pure 
configurations, there exist pictures where mixing is  included and the 
$f_0(980)$ incorporates both a $q\bar q$ and a $K\bar K$ or a glueball 
component~\cite{Tornqvist:1995kr,Anisovich:2002ij}. 

Note, that the question on at least the dominant structure of the 
$f_0(980)$ meson still remains open. Several comprehensive theoretical 
studies give completely opposing conclusions. {\it E.g.}, the unitarised 
meson model of~\cite{vanBeveren:1998qe} predicts two complete scalar 
meson nonets, where the $f_0(980)$ is considered as the $s\bar s$ ground 
state. ref.~\cite{Tornqvist:1995kr} also describes the $f_0(980)$ as 
a $q\bar q$ state but with a large $K\bar K$ component due to the 
proximity to the $K\bar K$ threshold. That is the meson spends most of 
its time in the virtual $K\bar K$ state. 
The analyses of~\cite{Anisovich:2001ay,Anisovich:2005kt} favor the 
$q\bar q$ interpretation of the $f_0(980)$. In particular, 
ref.~\cite{Anisovich:2005kt} predicted two poles close to the $K\bar K$ 
threshold, which suggests that the $f_0(980)$ is a $q\bar q$ state with 
a large $s\bar s$ component. 

The resonance structure of the $f_0(980)$ was analyzed 
in~\cite{Morgan:1993td} by using $J/\psi$ decay data. The $f_0$ was 
found to be a conventional Breit-Wigner structure. But the data are also 
compatible with one pole near threshold, which can be identified with 
a kaon bound state as mentioned in~\cite{Janssen:1994wn}.  
However, despite this controversial and detailed discussion concerning 
the $f_0$ structure, the $K\bar K$ bound state configuration 
seems to be the dominant contribution~\cite{Baru:2003qq,Hanhart:2007wa}.  

In the present paper, the $f_0(980)$ is considered as a pure 
$K\bar K$ molecule in a phenomenological Lagrangian approach. 
The coupling between the $f_0(980)$ meson and its constituents 
($K$ and $\bar K$ mesons) is described by the strong interaction  
Lagrangian. The corresponding coupling constant is determined by the  
compositeness condition $Z=0$~\cite{Weinberg:1962hj,Efimov:1993ei},  
which implies that the renormalization constant of the hadron 
wave function is set equal to zero. This condition was first applied 
in order to study the deuteron as a bound state of proton and 
neutron~\cite{Weinberg:1962hj}. Later this method was successfully 
applied to low-energy hadron phenomenology. It provides the basic 
equation for the covariant description of mesons and baryons as 
composite objects of light and heavy constituent quarks, as well 
as for glueballs which are bound states of gluons (see {\it e.g.} 
discussion in refs.~\cite{Efimov:1993ei,Efimov:1987sa,%
Ivanov:1997ug,Anikin:1995cf,Faessler:2003yf,Giacosa:2007bs}). 
Recently the compositeness condition was also used to study
the light scalar mesons $a_0$ and $f_0$ as $K \bar K$
molecules~\cite{Baru:2003qq,Hanhart:2007wa}. Here, in a first step, 
we apply our formalism to the study of the strong $f_0(980) \to \pi\pi$ 
and electromagnetic $f_0(980) \to \gamma\gamma$ decays.
In particular, previous determinations of the radiative decay width of the
$f_0 (980)$, for example when applying the quasi-static approximation,
suffer from large uncertainties due to a possible violation of local gauge
invariance (for a discussion on this issue see for  
example~\cite{Hanhart:2007wa}). In the present approach such uncertainties  
are avoided since we use a fully covariant and gauge invariant formalism. 

In the 
future we plan to extend the application to the $a_0(980)$ and 
investigate a possible $f_0(980)-a_0(980)$ mixture. Recently our 
Lagrangian approach, based on the compositeness condition, was 
successfully applied to the study of the $D_{s0}^\ast(2317)$ and 
$D_{s1}(2460)$ mesons considered as $(D K)$ and $(D^\ast K)$ 
molecules, respectively~\cite{Faessler:2007gv}.   
In the context of this formalism the strong, electromagnetic and 
weak decay properties of these states have been evaluated.  

In the present paper we proceed as follows. First, in sect.~\ref{sec:model}, 
we discuss the basic notions of our approach. We derive the 
phenomenological mesonic Lagrangian including photons for the 
treatment of the decay properties of the $f_0(980)$ meson as 
a $K\bar K$ bound state. Then, in sect.~\ref{sec:em}, we discuss 
the electromagnetic decay $f_0(980) \to \gamma\gamma$ with the 
associated diagrams and matrix elements. Special attention will 
be paid to the proof of electromagnetic gauge invariance.   
In sect.~\ref{sec:strong} we turn to the strong 
decay $f_0 \to \pi\pi$. Numerical results are discussed in 
sect.~\ref{sec:numerical}, followed by a short summary 
of our results in sect.~\ref{sec:summary}. 

\section{Approach}\label{sec:model}

In this section we derive and present the formalism for the 
study of the $f_0(980)$ meson as a hadronic molecule -
a bound state of $K$ and $\bar K$ mesons. 
This means that in our approach the $f_0(980)$ does 
not decay into a $K \bar K$ pair. 
Our framework is based on an interaction
Lagrangian describing the coupling between the $f_0(980)$ 
meson and its constituents as
\eq\label{f0KK_int} 
{\cal L}_{f_0K\bar K}(x) = g_{f_0K\bar K} \, f_0(x) 
\int dy \, \Phi(y^2) \, K^\dagger(x_-) \, K(x_+) \,, 
\en 
where $x_\pm = x \pm y/2$, $K=(K^+,K^0)$ and $K^\dagger =(K^-,\bar K^0)$ 
are the doublets of kaons and antikaons, $g_{f_0K\bar K}$ is the 
$f_0 K \bar K$ coupling constant. In particular, the assumed molecular 
structure of the $f_0(980)$ is in terms of particle content of the form
\eq
|f_0(980) \ra \, = \, \frac{1}{\sqrt{2}} \, 
( \, |K^+ K^-\ra + |K^0 \bar K^0\ra \, ) \,. 
\en 
The correlation function $\Phi$ in eq.~(\ref{f0KK_int}) 
characterizes the finite size of the $f_0(980)$ meson as a $(K\bar K)$ 
bound state and depends on the relative Jacobi coordinate $y$ and
 the center of mass (CM) coordinate $x$.
The local limit corresponds to the substitution of
$\Phi$ by the Dirac delta-function:
$\Phi(y^2) \to \delta^4 (y)$.
The Fourier transform of the correlation function
reads
\eq
\Phi(y^2) \, = \, \int\!\frac{d^4p}{(2\pi)^4}  \,
e^{-ip y} \, {\widetilde{\Phi}}(-p^2) \,.
\en
Any choice for $\widetilde\Phi$ is appropriate
as long as it falls off sufficiently fast in the ultraviolet region
of Euclidean space to render the Feynman diagrams ultraviolet finite.
We employ the Gaussian form
\eq\label{Gauss_CF}
\widetilde\Phi(p_E^2)
\doteq \exp( - p_E^2/\Lambda^2)
\en
for the vertex function, where $p_{E}$ is the
Euclidean Jacobi momentum. Here $\Lambda$ 
is a size constant, which parametrizes the distribution of 
kaons inside the $f_0$ molecule. 

The $f_0 K \bar K$ coupling constant $g_{f_0 K \bar K}$ 
is determined by the compositeness
condition~\cite{Weinberg:1962hj,Efimov:1993ei}, 
which implies that the renormalization constant of the hadron
wave function is set equal to zero:
\eq\label{Zf0} 
Z_{f_0} = 1 - \Sigma^\prime_{f_0}(M_{f_0}^2) = 0 \,.
\en
Here
$\Sigma^\prime_{f_0}(M_{f_0}^2)  = 
g_{f_0 K \bar K}^2 \Pi^\prime_{f_0}(M_{f_0}^2) $
is the derivative of the $f_0(980)$ meson mass operator described 
by the diagram in fig.~\ref{fig:massop} (details of the calculation are given 
in Appendix A). As we already stressed in the Introduction, this condition has been widely applied.   
 
In order to calculate the strong $f_0(980) \to \pi\pi$ and 
electromagnetic $f_0(980) \to  \gamma\gamma$ decays we need 
to specify the phenomenological Lagrangian which generates 
the contributing meson-loop diagrams. The lowest-order Lagrangian 
${\cal L}$, formulated in terms of the scalar meson $f_0(980)$, 
pseudoscalar ($\pi$, $K$, $\cdots$), vector ($K^\ast$, $\cdots$) 
mesons and photon fields, is given by: 
\eq 
{\cal L}(x) = {\cal L}_{f_0}(x) + {\cal L}_{f_0K\bar K}^{\rm GI}(x) 
+ {\cal L}_U(x) + {\cal L}_{W}(x)\,, 
\en 
where 
\eq 
{\cal L}_{f_0}(x) = - \frac{1}{2} f_0(x) (\Box + M_{f_0}^2) f_0(x) 
\en 
is the free Lagrangian of the $f_0$ meson with 
$\Box = \partial_\mu \partial^\mu$.\newline
${\cal L}_{f_0K\bar K}^{GI}$ is the gauge-invariant form of the 
$f_0K\bar K$ interaction Lagrangian, which {\it i.e.} 
includes photons via the path integral 
$I(x,y,P) = \int\limits_y^x dz_\mu A^\mu(z)$ with
\eq\label{L_str_em} 
\hspace*{-.4cm}
&&{\cal L}_{f_0K\bar K}^{\rm GI}(x)=g_{f_0K\bar K} f_0(x) \int dy 
\,\Phi(y)\nonumber\\
\hspace*{-.4cm}
&&\times\biggl[ e^{-ieI(x_+,x_-)} \, K^+(x_+) \, K^-(x_-) + \, 
K^0(x_+) \, \bar K^0(x_-) \biggr].    
\en 
The term ${\cal L}_U$, written as 
\eq 
{\cal L}_U(x) = \frac{F^2}{4} \la \, D_\mu U(x) D^\mu U^\dagger(x) + 
\chi U^\dagger(x) + \chi^\dagger U(x) \, \ra 
\en 
is the Lagrangian of second-order chiral perturbation theory 
(ChPT)~\cite{Weinberg:1978kz,Gasser:1983yg} in the 
three-flavor meson sector~\cite{Gasser:1984gg} and the Lagrangian
\eq\label{LW}  
{\cal L}_{W}(x) 
&=& \, \la \, - \frac{1}{2} \nabla^\sigma W_{\sigma\mu} 
\nabla_\nu W^{\nu\mu} + \frac{1}{4} M_V^2 W_{\mu\nu} W^{\mu\nu}\nonumber\\
&+& \frac{i G_V}{\sqrt{2}} W_{\mu\nu} u^\mu u^\nu \, \ra  
\en  
involves vector mesons in the tensorial 
representation~\cite{Gasser:1983yg,Ecker:1988te,Ecker:1989yg}.  
The symbols $\la \,\, \ra$ and $[ \,\, ]$ 
occurring in above formulas denote the trace over flavor
matrices and the commutator, respectively.

Here we use the standard notations of ChPT.   
The fields of pseudoscalar mesons are collected in the chiral matrix 
$U = u^2 = \exp(i\sum_i \phi_i \lambda_i/F)$ with $F = F_\pi = 92.4$ MeV 
being the leptonic decay constant; 
$D_\mu$ is the covariant derivative acting on the chiral 
field: 
\eq 
D_\mu U = \partial_\mu U + i [U, Q] A_\mu + \ldots \, 
\en  
with $Q$ being the charge matrix of quarks and $A_\mu$ is the 
electromagnetic field; $u_\mu = i u^\dagger D_\mu U u^\dagger$ 
is the chiral vielbein and $\chi = 2 B {\cal M} + \cdots$; $B$ 
is the quark vacuum condensate parameter 
$B = - \la 0|\bar u u|0 \ra/F^2 = - \la 0|\bar d d|0 \ra/F^2$;  
${\cal M} = {\rm diag}\{\hat m, \hat m, m_s\}$ is 
the mass matrix of current quarks with $\hat m = (m_u + m_d)/2$. 
We work in the isospin limit and rely on the standard picture of 
chiral symmetry breaking~($B \gg F$). In the leading order of 
the chiral expansion the masses of pions and kaons are given by
$M_{\pi}^2=2 \hat m B, \hspace*{.2cm} M_{K}^2=(\hat m + m_s) B\,.$   
$W_{\mu\nu} = - W_{\nu\mu}  = ( \sum_i W_i \lambda_i )_{\mu\nu}/\sqrt{2}$  
is the octet of vector fields written in the tensor 
representation in the terms of antisymmetric tensor 
fields~\cite{Gasser:1983yg,Ecker:1988te,Ecker:1989yg}; 
$\nabla^\mu$ is the covariant derivative acting on vector fields 
(in our case we restrict $\nabla^\mu$ to the simple derivative 
$\partial^\mu$); $G_V$ is the coupling related to the decay 
constant of a vector meson into two pseudoscalars. With the use 
of low-energy theorems it can be expressed through the leptonic 
decay constant $F$ as~\cite{Kawarabayashi:1966kd}: $G_V = F/\sqrt{2}$. 
Finally we comment on the Lagrangian ${\cal L}_{f_0K\bar K}^{\rm GI}$.  
As is discussed in detail in refs.~\cite{Anikin:1995cf,Faessler:2003yf}  
the presence of vertex form factors in the interaction Lagrangian 
[like the strong interaction Lagrangian describing the coupling of 
$f_0(980)$ to its constituents, see eq.~(\ref{f0KK_int})] requires 
special care in establishing gauge invariance. 
One of the possibilities is provided by a modification of the charged 
fields, which are multiplied by an exponential containing the 
electromagnetic field. This procedure was suggested 
in~\cite{Mandelstam:1962mi} and applied in ref.~\cite{Terning:1991yt} 
and in refs.~\cite{Anikin:1995cf,Faessler:2003yf,Faessler:2007gv}. 
In our case the fields of charged kaons are modified as
\eq\label{L_str_gauging}
K^\pm(y) \,\rightarrow \, e^{\mp i I(y,x,P)} \, K^\pm(y) \,. 
\en
By using the substitution (\ref{L_str_gauging}) we obtain the
strong Lagrangian ${\cal L}_{f_0K\bar K}^{\rm GI}$, 
which in addition obeys electromagnetic gauge invariance. 
Note, that for the derivative of the path integral 
$I(y,x,P)=\int\limits_x^y\,dz_\mu A^\mu(z)$  we use the
path-independent prescription suggested in ref.~\cite{Mandelstam:1962mi}
\eq\label{path1}
&&\lim\limits_{dx^\mu \to 0} dx^\mu
\frac{\partial}{\partial x^\mu} I(x,y,P) \nonumber\\
&&= \,
\lim\limits_{dx^\mu \to 0} [ I(x + dx,y,P^\prime) - I(x,y,P) ] \,,
\en
where the path $P^\prime$ is obtained from $P$ by shifting the end-point $x$
by $dx$. Definition~(\ref{path1})
leads to the key rule
\begin{eqnarray}\label{path2}
\frac{\partial}{\partial x^\mu} I(x,y,P) = A_\mu(x) \,,
\end{eqnarray}
which in turn states that the derivative of the path integral $I(x,y,P)$ does
not depend on the path P originally used in the definition. The non-minimal
prescription (\ref{L_str_gauging}) is therefore completely equivalent to the
minimal substitution (see {\it e.g.} discussion in ref.~\cite{Terning:1991yt}).   
Expanding the gauge exponential in eq.~(\ref{L_str_em}) up to second order 
in the electromagnetic field $A_\mu$ we generate nonlocal vertices with 
a single [fig.~\ref{fig:strong}(a)] and two [fig.~\ref{fig:strong}(b)] photons attached. 
The derivation of the Feynman rules for such vertices 
is discussed in detail in Appendix B. 

Now we turn to the discussion of the diagrams contributing to the strong 
$f_0(980) \to \pi\pi$ and electromagnetic $f_0(980) \to \gamma\gamma$  
decays. The diagrams which describe the two pion decay $f_0(980) \to \pi\pi$ 
are shown in fig.~\ref{fig:em}. The two-point diagram of fig.~\ref{fig:em}(a) is generated by 
the $K \bar K \pi\pi$ contact interaction, whereas the three-point diagram 
of fig.~\ref{fig:em}(b) includes an exchange of the vector meson $K^\ast$ 
in order to take into account rescattering effects. 
The diagrams describing the electromagnetic $f_0(980) \to \gamma\gamma$ 
decay are shown in fig.~\ref{fig:vertex}. In addition to the diagrams of fig.~\ref{fig:vertex}(a) 
and 4(b), which are generated by the coupling of charged kaons to photons, 
the restoration of electromagnetic gauge invariance requires the inclusion 
of the diagrams displayed in figs.~\ref{fig:vertex}(c)-\ref{fig:vertex}(e). These additional graphs are 
generated by the vertices shown in figs.~\ref{fig:strong}(a) and ~\ref{fig:strong}(b). 

Meson loop diagrams are evaluated by using the free meson propagators 
of $K$ and $K^\ast$. The $K$-meson propagator is given by 
\eq
i \, S_K(x-y)& =& \langle 0 | T \, K(x) \, K^\dagger(y)  | 0 \rangle
\nonumber\\ 
&=&\int\frac{d^4k}{(2\pi)^4i} \, e^{-ik(x-y)} \ \tilde S_K(k) \,,
\en
where 
\eq
\tilde S_K(k) = \frac{1}{M_K^2 - k^2 - i\epsilon}\,. 
\en 
For the $K^\ast$ meson we use the tensorial representation 
of the chiral Lagrangian (\ref{LW}) with the propagator
\eq\label{Kstar_tensor} 
S_{K^\ast; \mu\nu,\alpha\beta}^W(x-y)&=&
\la 0|T \, K_{\mu\nu}^{\ast}(x) K_{\alpha\beta}^{\ast \, \dagger}(y) |0 \ra 
\nonumber\\
&=&- \frac{1}{M_{K^\ast}^2}\int\frac{d^4k}{(2\pi)^4 i}
\frac{e^{ik\cdot(x-y)}}{M_{K^\ast}^2-k^2-i\epsilon} \nonumber\\ 
&\times&[g_{\mu\alpha}g_{\nu\beta}(M_{K^\ast}^2-k^2) 
+g_{\mu\alpha}k_{\nu}k_{\beta}\nonumber\\
&-&g_{\mu\beta}k_{\nu}k_{\alpha}
-(\mu\leftrightarrow\nu)]\,.  
\en 
In the numerical calculations we restrict to the isospin limit 
and identify the meson masses of the iso-multiplets
with the masses of the charged partners~\cite{Yao:2006px}
\begin{eqnarray}
M_\pi &\equiv& M_{\pi^\pm} = 139.57018 \ {\rm MeV}\,,
\nonumber\\
M_K &\equiv& M_{K^\pm}= 493.677 \ {\rm MeV}\,,\\ 
M_{K^\ast} &\equiv& M_{K^{\ast \, \pm}} 
= 891.66 \ {\rm MeV}\,. 
\nonumber 
\end{eqnarray}
Following refs.~\cite{Baru:2003qq,Hanhart:2007wa} we 
write the $f_0(980)$-mass in the form 
\eq 
M_{f_0} = 2 M_K - \epsilon\,,   
\en 
where $\epsilon$ represents the binding energy. Further on we will 
discuss the dependence of observables on $\epsilon$, considered as 
a free parameter.

\section{The electromagnetic decay} \label{sec:em}

In this section we consider the electromagnetic decay $f_0\to\gamma\gamma$, 
which proceeds via the charged kaon loop and 
where all appropriate diagrams are pictured in fig.~\ref{fig:vertex}. 
We derive the transition amplitude in a manifest gauge-invariant way 
and finally deduce the form factors.
In the first part, we consider the local case, which corresponds to 
a vertex function with 
$\lim\limits_{\Lambda\rightarrow\infty}\widetilde\Phi(-k^2)=1$ 
in the phenomenological Lagrangian (\ref{L_str_em}).  
After that we proceed with the nonlocal case keeping the full form of 
$\widetilde\Phi(-k^2)$ in (\ref{L_str_em}), {\it i.e.}  
include finite-size effects. As mentioned before, in the local 
approximation the two-photon decay amplitude is generated by the two 
diagrams of figs.~\ref{fig:vertex}(a) and \ref{fig:vertex}(b). In the nonlocal case we have to include 
the three additional diagrams of figs.~\ref{fig:vertex}(c)-\ref{fig:vertex}(e) to restore the full 
electromagnetic gauge invariance. For the evaluation of the diagrams 
we follow the procedure developed in 
refs.~\cite{Anikin:1995cf,Faessler:2003yf,Faessler:2007gv,Giacosa:2007bs}. 
{\it E.g.} in ref.~\cite{Faessler:2003yf} we present a detailed analysis 
of the two-photon decay of the light $\sigma$-meson, which is similar 
to the present case. The contribution of each single diagram is not 
gauge invariant by itself, but the total sum is invariant. 
Therefore, the calculation of the matrix elements can be simplified by 
separating each diagram into a gauge invariant part $I^{\mu\nu}_\perp$ 
and a remainder, denoted by $\delta I^{\mu\nu}$, as 
\eq\label{Imunu} 
I^{\mu\nu}=I_\perp^{\mu\nu}+\delta I^{\mu\nu} \,.  
\en 
This separation can be achieved in the following 
manner.
For the $\gamma$-matrices (which only appear in fermionic loops) 
and vectors with open Lorentz index $\mu$, coinciding with the index 
of the photon polarization vector $\epsilon_\mu^{(\lambda)}$, one can 
use the representation (see {\it e.g.} the discussion in 
refs.~\cite{Anikin:1995cf,Faessler:2003yf,Faessler:2007gv}): 
\eq 
\gamma_\mu = \gamma_\mu^\perp + q_\mu \frac{\not\! q}{q^2} \,, 
\hspace*{1cm}  
p_\mu = p_\mu^\perp + q_\mu \frac{pq}{q^2} \,. 
\en  
The sum of all remainder terms cancels (see Appendices C and D). 
Therefore, the resulting matrix element is only given by the first part 
in the r.h.s. of~(\ref{Imunu}), from which we can extract the form factors. 

The $f_0(980) \to \gamma \gamma$ transition matrix element 
${\cal M}_{f_0}^{\mu\nu}$ can be written in terms of the Lorentz 
tensors $b^{\mu\nu}$ and $c^{\mu\nu}$ as
\eq 
{\cal M}^{\mu\nu}_{f_0} &=& e^2 \biggl\{ 
F_{f_0\gamma\gamma}(p^2,q_1^2,q_2^2)
\,b^{\mu\nu}\nonumber\\
&+&G_{f_0\gamma\gamma}(p^2,q_1^2,q_2^2)\,c^{\mu\nu} 
\biggr\}\,.
\en 
The tensor structures are given by
\eq
b^{\mu\nu}&=&g^{\mu\nu} (q_1q_2) - q_1^\nu q_2^\mu\,,\\
c^{\mu\nu}&=&g^{\mu\nu}q_1^2q_2^2 + q_1^\mu q_2^\nu(q_1q_2) 
- q_1^\mu q_1^\nu q_2^2 - q_2^\mu q_2^\nu q_1^2 \,,
\en 
where $q_1$ and $q_2$ refer to the four-momenta of the photons. 
Note, that the part of the matrix element containing the form factor 
$G_{f_0\gamma\gamma}(p^2,\,q_1^2,\,q_2^2)$ vanishes, when at least 
one of the photons is on-shell, since $c^{\mu\nu}=0$. \newline
The two photon decay width 
\eq
\Gamma(f_0\to\gamma\gamma) = 
\frac{\pi}{4}\alpha^2 M_{f_0}^3 g_{f_0\gamma\gamma}^2		
\label{eq: width}
\en
is expressed in terms of the coupling 
constant 
\eq\label{eq: couplconst}
g_{f_0\gamma\gamma}=F_{f_0\gamma\gamma}(M_{f_0}^2,\,0,\,0)\,,
\en 
which in turn is related to the form factor $F_{f_0\gamma\gamma}$.

\subsubsection{Local case} 

In the local approximation we only have to consider the diagrams of 
figs.~\ref{fig:vertex}(a) and \ref{fig:vertex}(b). Using the decomposition (\ref{Imunu}) one can show 
that the resulting matrix element is given by the 
gauge-invariant part of the diagram of fig.~\ref{fig:vertex}(a) only with   
\eq 
I_{\Delta, \perp}^{L \, \mu\nu}(q_1,\,q_2)& =& 
\int\,\frac{d^4k}{\pi^2i}\,(2k+q_1)_{\perp q_1}^\mu\,
(2k-q_2)_{\perp q_2}^\nu\nonumber\\
&\times&S_K\left(k+q_1\right)\,S_K\left(k-q_2\right)
\,S_K(k)\,,			
\label{eq: local}
\en
which is ultraviolet (UV) finite. An important point is that 
noninvariant parts of figs.~\ref{fig:em}(a) and \ref{fig:em}(b) cancel each other 
(see details in Appendix C). The evaluation of eq.~(\ref{eq: local}) 
yields (the explicit calculation is carried out in Appendix C): 
\eq 
F_\Delta^L(p^2,q_1^2,q_2^2) = \int\limits_0^1d^3\alpha 
\,\delta(1-\sum\limits_{i=0}^3\alpha_i)\frac{4\alpha_1\alpha_2}{D}\,, 
\label{eq: flocal}
\en
where $D = M_K^2 - p^2 \alpha_1\alpha_2 - q_1^2 \alpha_1\alpha_3 
- q_2^2 \alpha_2\alpha_3$. Later on, we also analyze the electromagnetic 
$f_0(980)$ form factors for different values of the photon virtuality 
in the Euclidean region by expressing 
$q_1^2$ and $q_2^2$ as $q_1^2 = - Q^2(1+\omega)/2$ and 
$q_2^2 = - Q^2(1-\omega)/2$. A similar 
analysis of the electromagnetic $\pi$ and $\sigma$-meson form factors 
was performed previously in refs.~\cite{Frank:1994gc} 
and~\cite{Ivanov:1997ug,Faessler:2003yf,Giacosa:2007bs}. 

The expression for the coupling constant $g_{f_0\gamma\gamma}$, 
resulting from \cite{Ivanov:1997ug}, is given by 
\eq 
g_{f_0\gamma\gamma} &=& \frac{g_{f_0K\bar K}}{8\pi^2} 
\, \int\limits_0^1d\alpha_1\int\limits_0^1d\alpha_2\,
\frac{4\alpha_1\alpha_2(1-\alpha_1)^2}{M_K^2-M_{f_0}^2\alpha_1
\alpha_2(1-\alpha_1)}\nonumber\\
& =& 
\frac{g_{f_0 K\bar K}}{8 \pi^2}R_{f_0\gamma\gamma}\,,
\en 
where 
\eq 
\vspace*{-3cm} 
R_{f_0\gamma\gamma} = \biggl(\frac{\arcsin\xi}{\xi}\biggr)^2 - 1\,, \;  
\xi = \frac{M_{f_0}}{2M_K} = 1 - \frac{\epsilon}{2M_K} \,. 
\en 
In the local limit (see Appendix A) the coupling constant $g_{f_0K\bar K}$ 
is expressed in the form
\eq 
\frac{1}{g_{f_0K\bar K}^2} = \frac{R_{f_0}}{8\pi^2 M_{f_0}^2}
\en 
with 
\eq 
R_{f_0} = \frac{\arcsin\xi}{\xi\sqrt{1-\xi^2}} - 1 \,.  
\en 
Finally, the expression for the $f_0\rightarrow\gamma\gamma$ decay 
width reads as follows
\eq\label{Gamma_fin}  
\Gamma(f_0\to\gamma\gamma) 
= \frac{\alpha^2}{8 \pi} \, \frac{R_{f_0\gamma\gamma}^2}{R_{f_0}}  
M_{f_0} \,, 
\en 
which agrees with the result of~\cite{Baru:2003qq,Hanhart:2007wa}.  

\subsubsection{Nonlocal case} 

Now we turn to the nonlocal interaction case and discuss the evaluation 
of the corresponding Feynman diagrams. 
In this case we have to incorporate all diagrams shown in fig.~\ref{fig:vertex}. 
In Appendix B we outline the derivation of the Feynman rules 
for the nonlocal vertices with a single [fig.~2(a)] and two [fig.~2(b)] 
photon lines attached; these nonlocal vertices have to be included to 
guarantee full gauge invariance both on the level of the Lagrangian and 
the matrix elements. 
For further details we refer to our earlier paper~\cite{Faessler:2003yf}.
In analogy to the local case, the form factors $F(p^2,q_1^2,q_2^2)$ 
and $G(p^2,q_1^2,q_2^2)$ can be extracted 
from the gauge invariant parts of the respective diagrams 
\eq 
I^{\mu\nu}(q_1,q_2) &\equiv& I^{\mu\nu}_{\perp}(q_1,q_2) \nonumber\\
&=&  
F(p^2,q_1^2,q_2^2) \, b^{\mu\nu} + G(p^2,q_1^2,q_2^2) \, c^{\mu\nu} \,.
\en 
The individual contributions are given by: 
\eq 
&&I^{\mu\nu}_{\triangle_\perp}(q_1,q_2) \nonumber\\ 
&&=\int\,\frac{d^4 k}{\pi^2i}\,\widetilde\Phi(-k^2)\, 
(2k+q_2)_{\perp q_1}^\mu\,(2k-q_1)_{\perp q_2}^\nu \nonumber\\ 
&&\times S_K\left(k+\frac p2\right)\,S_K\left(k-\frac p2\right)\, 
S_K\left(k+\frac q2\right)
\nonumber\\
&&\nonumber\\
&&= F_{\triangle}(p^2,q_1^2,q_2^2) b^{\mu\nu}+
    G_{\triangle}(p^2,q_1^2,q_2^2) c^{\mu\nu}
\label{eq: tri}\\  
&&I^{\mu\nu}_{\rm bub_\perp}(q_1,q_2)\nonumber\\ 
&&= -\,\int\frac{d^4 k}{\pi^2 i}\int\limits_0^1\!dt \,
\widetilde\Phi^\prime(- x(0,q_1))\, 
2k^\mu_{\perp q_1}k^\nu_{\perp \, q_2}\nonumber\\
&&\times\,S_K\left(k+\frac {q_2}{2}\right)\,S_K\left(k-\frac {q_2}{2}\right)
\nonumber\\
&&+(q_1\leftrightarrow q_2,\,\mu\leftrightarrow \nu)\nonumber\\
& &\nonumber\\
&&=  F_{\rm bub}(p^2,q_1^2,q_2^2) b^{\mu\nu} +
     G_{\rm bub}(p^2,q_1^2,q_2^2) c^{\mu\nu} 
\label{eq: bub} 
\en
and
\eq
&&I^{\mu\nu}_{\rm tad_\perp}(q_1,q_2) \nonumber\\
&&=\int\!\frac{d^4 k}{\pi^2 i} \, S_K(k)
\int\limits_0^1\!dt \, \Biggl( - \, \frac{c^{\mu\nu}}{4\,q_1^2\,q_2^2} \, 
\biggl( \widetilde\Phi^\prime(- x(0,p) )\nonumber\\ 
&&+\widetilde\Phi^\prime(- x(0,q) ) \biggr)+ t \int\limits_0^1\!dl\, 
\biggl(k+\frac{q_2}{2}\biggr)^\mu_{\perp \, q_1} 
k^\nu_{\perp q_2}\nonumber \\ 
&&\times\biggl( \, \widetilde\Phi^{\prime\prime}(- x(q_1,q_2)) + 
\widetilde\Phi^{\prime\prime}(- x(- q_1,q_2))\biggr) \Biggr)\nonumber\\
&&+(q_1\leftrightarrow q_2,\,\mu\leftrightarrow \nu)\nonumber\\
& &\nonumber\\
&&= F_{\rm tad_\perp}(p^2,\,q_1^2,\,q_2^2)b^{\mu\nu}+
        G_{\rm tad_\perp}(p^2,\,q_1^2,\,q_2^2)c^{\mu\nu}\,, 
\label{eq: tad} 
\en 
where 
\eq
x(q_1,q_2) \,& =& \,  k^2 + k t (l\,q_1+q_2)
            +\frac{t}{4}(l\,q_1^2+2\,l\,q_1q_2+q_2^2)\,, \nonumber\\
q \, &=& \, q_2 - q_1 \,. \label{eq: x}
\en
Here, eqs.~(\ref{eq: tri}), (\ref{eq: bub}) and (\ref{eq: tad}) correspond 
to the triangle diagrams of figs.~4(a) and 4(b), to the bubble 
diagrams of figs.~4(c) and 4(d), and to the tadpole diagram of fig.~4(e). 
The evaluation of these structure integrals is performed in 
Appendix D without referring to a specific functional form of the 
vertex function $\widetilde\Phi(-k^2)$. 
Then the decay width for $f_0(980) \to \gamma\gamma$ 
is calculated in analogy to the local case. 
We discuss the numerical results in sect.~\ref{sec:numerical}. 

\section{The strong decay} \label{sec:strong}

In this section we discuss the features of the $f_0(980) \to \pi\pi$ decay.  
The two-pion decay width is determined by the expression  
\eq 
\Gamma_{f_0\pi\pi} &=& \Gamma_{f_0\pi^+\pi^-}+\Gamma_{f_0\pi^0\pi^0} = 
\frac{3}{2}\Gamma_{f_0\pi^+\pi^-}\nonumber\\
&=& \frac{3}{32\pi}\frac{g_{f_0\pi\pi}^2}{M_{f_0}}
\sqrt{1-\frac{4M_{\pi}^2}{M_{f_0}^2}}\,, \label{eq: widthstr}
\en 
where the coupling constant $g_{f_0\pi\pi}$ is given by
\eq 
g_{f_0\pi\pi} = g_{f_0\pi^+\pi^-}=2g_{f_0\pi^0\pi^0}=G(M_{f_0}^2,M_\pi^2,M_{\pi}^2)
\en 
defined by the effective Lagrangian ${\cal L}_{f_0\pi\pi}=\frac12g_{f_0\pi\pi}f_0\vec \pi^2$.
Here, $G(p^2,q_1^2,q_2^2)$ is the structure integral of the 
$f_0 \to \pi \pi$ transition, which is conventionally split into 
the two terms $G^{(a)}(p^2,\,q_1^2,\,q_2^2)$ and 
$G^{(b)}(p^2,\,q_1^2,\,q_2^2)$. They refer to the 
contributions of the diagrams of figs.~3(a) and 3(b), 
respectively, with 
\eq 
\hspace*{-.3cm} 
G(p^2,q_1^2,q_2^2)&=& G^{(a)}(p^2,q_1^2,q_2^2) + 
                       G^{(b)}(p^2,q_1^2,q_2^2) \,, 
\en 
\eq 
\hspace*{-.8cm} 
G^{(a)}(p^2,q_1^2,q_2^2) &=& \frac{g_{f_0 K \bar K}}{3 \, F^2}   
\int\frac{d^4k}{(2\pi)^4 i} \, \widetilde\Phi(-k^2) \nonumber\\
&\times&\frac{M_K^2 + \frac{5}{4} p^2 - k^2}
{(M_K^2 - k^2_+) (M_K^2 - k^2_-)}\,, 
\en 
\eq 
\hspace*{-.10cm} 
&&G^{(b)}(p^2,q_1^2,q_2^2) = \frac{g_{f_0 K \bar K}}{2 \, F^2}   
\int\frac{d^4k}{(2\pi)^4 i} \, \widetilde\Phi(-k^2)
\nonumber\\
&&\times\Big\{\frac{(k^2_+ - q_1^2) (k_+ - q_1)(k_- -q_2) }{(M_K^2 - k^2_+) 
(M_K^2 - k^2_-) (M_{K^\ast}^2 - (k_+ - q_1)^2)}\nonumber\\
&&-\frac{(k_+ - q_1)^2 (k_+ + q_1) (k_- - q_2)}
{(M_K^2 - k^2_+) (M_K^2 - k^2_-) (M_{K^\ast}^2 - (k_+ - q_1)^2)}\Big\}\,, 
\label{Gf0pipi}
\en 
where $k_\pm = k \pm p/2$. Here we substitute $G_V = F/\sqrt{2}$ 
in the expression for $G^{(b)}(p^2,q_1^2,q_2^2)$. 

It is worth to note, that the sum of the diagrams of fig.~3 contributing 
to the coupling $g_{f_0\pi\pi}$ can be approximated by a single diagram - 
the triangle diagram of fig.~3(b) with the exchanged $K^\ast$ meson 
propagator in vectorial representation. In particular, we remind that 
the free Lagrangian of the $K^\ast$ meson in the vector representation is 
written in the form: 
\eq 
{\cal L}_V = - \frac{1}{2} K^{\ast\, \dagger}_{\mu\nu} K^{\ast \, \mu\nu} 
+ M_{K^\ast}^2 K_\mu^{\ast \, \dagger} K^{\ast \mu},
\en
where $K^\ast_{\mu\nu} = \partial_\mu K^\ast_\nu - \partial_\nu K^\ast_\mu$. 
Then for the sake of comparison between the two different representation 
it is convenient to write down the propagator in the vector representation 
as a $T$-product of $K^\ast_{\mu\nu}$: 
\begin{eqnarray}\label{Propagator_V} 
\hspace*{-.7cm} 
S_{K^\ast; \mu\nu,\alpha\beta}^V(x-y)&=&
\langle0|T \, K^\ast_{\mu\nu}(x) K^{\ast \, \dagger}_{\alpha\beta}(y) 
|0\rangle \nonumber\\
\hspace*{-.7cm}
&=& - \frac{1}{M_{K^\ast}^2}\int\frac{d^4k}{(2\pi)^4 i}
\frac{e^{ik\cdot(x-y)}}{M_{K^\ast}^2-k^2-i\epsilon} \nonumber\\ 
\hspace*{-.7cm}
&\times&[g_{\mu\alpha}k_{\nu}k_{\beta}
-g_{\mu\beta}k_{\nu}k_{\alpha}-(\mu\leftrightarrow\nu)] \,.
\end{eqnarray} 
As was stressed in ref.~\cite{Ecker:1989yg}, the propagators 
$S_{K^\ast; \mu\nu,\alpha\beta}^V$ and 
$S_{K^\ast; \mu\nu,\alpha\beta}^W$ 
differ by the contact term contained in the tensorial propagator: 
\eq\label{relation_VW} 
S_{K^\ast; \mu\nu,\alpha\beta}^W(x) 
&=& S_{K^\ast; \mu\nu,\alpha\beta}^V(x)\nonumber\\  
&+& \frac{i}{M_{K^\ast}^2}
[g_{\mu\alpha}g_{\nu\beta} - g_{\mu\beta}g_{\nu\alpha}]  \, \delta^4(x) \,. 
\en 
Using identity (\ref{relation_VW}) one can show that the contribution of 
the diagram fig.~3(b) in tensorial representation is given by the sum 
of the graph of fig.~3(b) in vectorial representation 
plus a graph, which is diagrammatically described by fig.~3(a), but has 
opposite sign and a different numerator in comparison to the structure 
integral $G^{(a)}(p^2,q_1^2,q_2^2)$. Latter graph results from the contact 
term contained in the propagator (\ref{relation_VW}), leading to the 
collapse of the $K^\ast$ line in fig.~3(b) to a point. In other words, 
the sum of diagrams of fig.~3 in tensorial representation generates 
a leading term which corresponds to the diagram of fig.~3(b) in vectorial 
representation. In addition we obtain a term resulting effectively from 
the difference of two graphs of the type fig.~3(a), but with different 
numerators in the expression. Numerically it is found that in the last 
term these two contributions almost compensate each other. Therefore, 
we effectively obtain a contribution of the graph of fig.~3(b) only, 
but now with the $K^\ast$ meson propagator written in 
vectorial representation.

As an alternative we also present a technique how to include non-perturbative pion-pion interaction near the $K\bar K$ threshold  by following the method of a chiral unitarity approach. By defining an effective coupling $\widetilde g_{K\bar K\pi\pi}$ between the intermediate kaons and the $\pi\pi$ pair we can write down the phenomenological Lagrangian
$$
{\cal L}_{K\bar K\pi\pi}=\widetilde g_{K\bar K\pi\pi}\bar KK\vec\pi\cdot\vec\pi\,.
$$
Within this approach the $f_0\pi\pi$ coupling and consequently the width is defined in terms of this effective coupling constant
\begin{eqnarray*}
&&g_{f_0\pi\pi}=\frac{g_{f_0K\bar K}\widetilde g_{K\bar K\pi\pi}}{(4\pi)^2}\\
&&\times\int\frac{d^4k}{\pi^2 i}\,\widetilde\Phi(-k^2)\frac{1}{M_K^2-\big(k+\frac p2\big)^2}\frac{1}{M_K^2-\big(k-\frac p2\big)^2}\,.
\end{eqnarray*}
A full analysis of this ansatz goes beyond the scope of the present paper. However, by using the two-pion decay width $\Gamma(f_0\to\pi\pi)$=19.5 MeV derived in \cite{Oller:1998hw} within the framework of a chiral unitarity approach, we can estimate the effective coupling $\widetilde g_{K\bar K\vec\pi\vec\pi}$ around 10 GeV.

\section{Numerical analysis}\label{sec:numerical}

\subsection{Coupling constant ${\mathbf g_{f_0K \bar K}}$} 

First we present our results for the $g_{f_0 K \bar K}$ coupling 
constant, both in the local and nonlocal case. In figs.~5 and 6 we 
demonstrate the sensitivity of the $g_{f_0 K \bar K}$ coupling on 
variations of the free parameters. In fig.~5 we give $g_{f_0 K \bar K}$ 
as a function of the binding energy $\epsilon$ for the local case. 
In fig.~6 $g_{f_0 K \bar K}$ is drawn as a function of two parameters, 
$\epsilon$ and the size parameter $\Lambda$ of the vertex function. 
For typical values of $\epsilon = 2 M_K -980$ MeV $\simeq 7.4$ MeV 
(corresponding to $M_{f_0} = 980$ MeV) and $\Lambda = 1$ GeV the results 
for the coupling constant are
\eq 
g_{f_0 K \bar K} = 2.90 \ {\rm GeV}  \ \ \ (\rm{local \ case}) 
\en 
and 
\eq 
g_{f_0 K \bar K} = 3.09 \ {\rm GeV}  \ \ \ (\rm{nonlocal \ case}) \,.  
\en 

\subsection{The electromagnetic decay}

In figs.~7 and 8 we present our results for the coupling constant 
$g_{f_0\gamma\gamma}$ for the local and the nonlocal case. 
In particular, in fig.~7 we consider the local limit and draw 
$g_{f_0\gamma\gamma}$ as a function of the binding energy $\epsilon$. 
The results for $g_{f_0\gamma\gamma}$ in the nonlocal case in dependence 
on the two parameters $\epsilon$ and $\Lambda$ is presented in fig.~8. 
For the typical values of $\epsilon = 7.4$ MeV 
and $\Lambda = 1$ GeV we obtain both for the coupling constant 
$g_{f_0\gamma\gamma}$ and the decay width $\Gamma(f_0\to\gamma\gamma)$ 
the results
\eq\label{Gf0gg} 
g_{f_0\gamma\gamma} &=& 0.086 \ {\rm GeV}^{-1}
\,,\nonumber\\
\Gamma(f_0\to\gamma\gamma) &=& 0.29 \ {\rm keV} \ \ \ (\rm{local \ case}) 
\en 
and 
\eq 
g_{f_0\gamma\gamma} &=& 0.079 \ {\rm GeV}^{-1}\,,\nonumber\\
\Gamma(f_0\to\gamma\gamma) &=& 0.25 \ {\rm keV}  
\ \ \ (\rm{nonlocal \ case}) \,.  
\en 
These results are in very good agreement with the new average 
$\Gamma(f_0\to\gamma\gamma)=0.29^{+0.07}_{-0.09}$ keV quoted 
by PDG 2007~\cite{Yao:2006px} and recent data presented by the
Belle Collaboration~\cite{Mori:2006jj}: $\Gamma(f_0\to\gamma\gamma) = 
0.205^{+ 0.095 + 0.147}_{- 0.083 - 0.117}$ keV. 
Our results also agree with previous results 
by the Crystal Ball Collaboration \cite{Marsiske:1990hx} 
($\Gamma_{\gamma\gamma}=0.31\pm0.14\pm0.09$ keV) and by 
MARK II at SLAC~\cite{Boyer:1990vu} 
($\Gamma_{\gamma\gamma}=0.29\pm0.07\pm0.12$ keV). 

In table~\ref{tab:em} we compare our predictions~(\ref{Gf0gg}) for the two-photon 
decay width of the $f_0$ meson with other 
theoretical approaches~\cite{Scadron:2003yg,Anisovich:2001zp,%
Oller:1997yg,Hanhart:2007wa,Efimov:1993ei,Achasov:1981kh,Schumacher:2006cy}. 
Quark-antiquark models predict similar results for 
$\Gamma(f_0\to\gamma\gamma)$. For example, predictions in the 
quarkonium interpretation range from 
$\Gamma(f_0\rightarrow\gamma\gamma)$=0.24 keV~\cite{Efimov:1993ei} 
to values of 0.33 keV \cite{Schumacher:2006cy} with an intermediate 
result of $0.28^{+0.09}_{-0.13}$ keV \cite{Anisovich:2001zp}. 
The result of the four-quark model \cite{Achasov:1981kh} with 
$\Gamma(f_0\rightarrow\gamma\gamma)$=0.27 keV lies in the range 
set by our local and nonlocal predictions. 

Previous determinations in the context of hadronic molecule 
interpretations of $\Gamma(f_0\rightarrow\gamma\gamma)$ = 
0.20 keV~\cite{Oller:1997yg} and $0.22\pm 0.07$~\cite{Hanhart:2007wa} 
lie at the lower side of our results. The difference in results can to 
some extent be explained by using different values for the meson masses 
({\it e.g.} as in ref.~\cite{Hanhart:2007wa}). 

At this level, present data on $\Gamma(f_0\rightarrow \gamma\gamma)$ cannot serve to uniquely deduce the dominant configuration of the $f_0(980)$. Furthermore, a recent amplitude analysis \cite{Pennington:2008xd} involving high statistics data by BELLE \cite{Mori:2006jj} allows for values for $\Gamma(f_0\rightarrow \gamma\gamma)$ from 0.10 to 0.54 keV. Therefore, a more precise determination of the width, narrowing down the region of the possible values, would affect the determination of the dominant configuration. 

For the $f_0\rightarrow\gamma\gamma$ decay properties 
finite size effects only play a role at the level of about 10\% 
(for the decay width) when both photons are on-shell (in this point 
we completely agree with the conclusions of ref.~\cite{Hanhart:2007wa}). 
The finite size effects become essential for a nontrivial virtuality 
of one of the photons. The same qualitative conclusion about the 
importance of finite size effects was known before from the analysis 
of the electromagnetic transition form factors of the $\pi$- and the 
light $\sigma$-meson (see~refs.~\cite{Frank:1994gc} 
and~\cite{Ivanov:1997ug,Faessler:2003yf}). In particular, it was shown 
(see {\it e.g.} discussion in refs.~\cite{Ivanov:1997ug,Faessler:2003yf}) 
that a local coupling between the $\pi$- or the light $\sigma$-meson and 
their constituent quarks leads to the wrong asymptotics, that is 
$\ln^2(Q^2/m^2)/(2Q^2)$, of the transition form factor, 
when one of the photons is on-shell while the other one has 
an Euclidean off-shell momentum squared. Such an asymptotics is in 
contradiction to the QCD-prediction for the $\pi^0\gamma\gamma^\ast$ 
form factor of $1/Q^2$~\cite{Brodsky:1981rp}. A similar situation holds 
for the electromagnetic $f_0$ form factor.
In fig.~9 we indicate the form factor $F_{f_0\gamma\gamma^\ast}(Q^2) = 
F_{f_0\gamma\gamma}(M_{f_0}^2,-Q^2,\,0)$
for the transition $f_0\rightarrow\gamma\gamma^\ast$ with a real and 
a virtual photon of Euclidean momentum squared $-Q^2$. To demonstrate 
the sensitivity of this form factor on finite-size effects we plot the 
result both for the local case and for the nonlocal vertex function 
with different values for $\Lambda = 0.7, 1$ and $1.3$ GeV. 
The additional bubble and tadpole diagrams of figs.~4(c)-(e), 
which result from the nonlocal interaction, only give a minor contribution 
to the dominant one of the triangle diagram. The curve corresponding to the local case lies considerably higher than 
those for the nonlocal case (including its asymptotic behavior at large 
values of~$Q^2$), even for moderate values of $Q^2$. Moreover, the 
electromagnetic form factor shows a sensitivity of about 20\% in the 
$Q^2$ range considered with respect to variations of $\Lambda$ or to 
the finite size of $f_0$. Clearly, an experimental determination of 
$F_{f_0\gamma\gamma^\ast}(Q^2)$ would help in possibly identifying the 
underlying structure of the $f_0(980)$. $F_{f_0\gamma\gamma^\ast}(Q^2)$  
can be approximated by a monopole function. Actually, this was also to 
be expected from the analysis of analogous form factors for the $\pi^0$ 
and the light $\sigma$ meson (see~\cite{Frank:1994gc} 
and~\cite{Ivanov:1997ug,Faessler:2003yf}) with: 
\eq\label{F_mon} 
F_{f_0\gamma\gamma^\ast}(Q^2) = 
\frac{F_{f_0\gamma\gamma^\ast}(0)}{1 + 
Q^2/\Lambda_{f_0\gamma\gamma^\ast}^2}\,,   
\en  
where $F_{f_0\gamma\gamma^\ast}(0) \equiv g_{f_0\gamma\gamma}$. 
The scale parameter $\Lambda_{f_0\gamma\gamma^\ast}$ can be related to 
the slope of the $F_{f_0\gamma\gamma^\ast}(Q^2)$ form factor defined as  
\eq\label{slope} 
\la r^2 \ra \, = \, - 6 \, 
\frac{F_{f_0\gamma\gamma^\ast}^\prime(0)}
{F_{f_0\gamma\gamma^\ast}(0)}\,,
\en 
where 
\eq 
F_{f_0\gamma\gamma^\ast}^\prime(0) = 
\frac{\partial F_{f_0\gamma\gamma^\ast}(Q^2)}
{\partial Q^2}\bigg|_{Q^2=0} \,. 
\en 
From eqs.~(\ref{F_mon}) and (\ref{slope}) we deduce the relation 
between $\la r^2 \ra$ and $\Lambda_{f_0\gamma\gamma^\ast}$ with 
\eq 
\la r^2 \ra = \frac{6}{\Lambda_{f_0\gamma\gamma^\ast}^2} \,.  
\en 
Therefore, the slope $\la r^2 \ra$ has a quite clear physical meaning - 
it is related to the charge distribution of the constituents $K^+$ and 
$K^-$ inside the $f_0$ molecule. 
Our prediction for $\la r^2 \ra$ at values of $\epsilon = 7.4$ MeV and 
$\Lambda = 1$ GeV is 
\eq 
\la r^2 \ra = 0.15 \ {\rm fm}^2 \,. 
\en 
It constraints the scale parameter to a value of 
$\Lambda_{f_0\gamma\gamma^\ast} = 1.26 $ GeV. 
Good agreement in the full $Q^2$ range with the exact result of the 
nonlocal case with parameters $\epsilon = 7.4$ MeV and $\Lambda = 1$ GeV 
is obtained with a monopole function of scale parameter 
$\Lambda_{f_0\gamma\gamma^\ast} 
\simeq 1.4$ GeV (for comparison it is also plotted in fig.~9), 
which is a bit larger than the value of 1.26 GeV deduced near $Q^2$=0. 

To improve the $F_{f_0\gamma\gamma^\ast}$ form factor we can also include finite size effects in the $K\bar K\gamma$ couplings, {\it i.e.} by replacing the local $K\bar K\gamma$ vertices by the electromagnetic kaon form factors 

$$F_{K\bar K\gamma}(Q^2)=\frac{1}{1+Q^2/\Lambda_{K\bar K\gamma}^2}\,.$$
The size parameter $\Lambda_{K\bar K\gamma}$ is related to the slope of the kaon electromagnetic form factor (\ref{slope}) which can be deduced from the charged kaon radius $\big<r_{K^\pm}^2\big>=\frac{6}{\Lambda_{K\bar K\gamma}}$ which we take from \cite{Yao:2006px}. The charge radius $r_{K^\pm}=0.56$ fm leads to a value of $\Lambda_{K\bar K\gamma}=0.863$ GeV. The additional form factors suppress the estimates for the coupling $g_{f_0\gamma\gamma^\ast}$ when the off-shell behavior of the photons is studied (see fig.~10). However, the finite size effects are still apparent. The decay properties for real photons are not affected by the monopole form factor since $F_{K\bar K\gamma}(Q^2)\big|_{Q^2=0}=1$.
\subsection{The strong decay}

The experimental results for the dominant decay process 
$f_0(980)\to\pi\pi$ are spread out in a large region. 
Consequently, PDG 2007~\cite{Yao:2006px} indicates a wide range 
for the total width from 40-100 MeV. 
Recently, the Belle Collaboration~\cite{Mori:2006jj} 
reported the result of 
$\Gamma(f_0\to\pi\pi)=51.3^{+20.8\,+13.2}_{-17.7\,-3.8}$ MeV. 

The coupling constant $g_{f_0\pi\pi}$ and, accordingly, the decay width 
(\ref{eq: widthstr}) depend sensitively on the cut-off parameter 
$\Lambda$, since without the correlation function $\widetilde\Phi(-k^2)$ 
the diagrams of fig.~3 are UV divergent. In fig.~11 we present our 
results for the coupling $g_{f_0\pi\pi}$ as a function of $\epsilon$ 
and $\Lambda$. Again the results for the coupling constant $g_{f_0\pi\pi}$ 
and the decay width $\Gamma(f_0\to\pi\pi)$ are given for the usual values 
of $\epsilon = 7.4$ MeV and $\Lambda = 1$ GeV as 
\eq\label{gf0pipi}
g_{f_0\pi\pi} = 1.53 \ {\rm GeV}  
\en 
and 
\eq 
\Gamma(f_0 \to \pi\pi) = 69 \ {\rm MeV}\,, 
\en 
consistent with current observation.

In table~\ref{tab:strong} we compare our results to data 
and results of other theoretical approaches. Predictions of $q \bar q$ 
models for $\Gamma(f_0\rightarrow\pi\pi)$ span a large range of values 
from 20 MeV~\cite{Efimov:1993ei} to 56 $-$ 58 MeV~\cite{Anisovich:2002ij}, 
which sensitively depend on the nonstrange flavor content. For the 
dynamically generated $f_0(980)$ of ref.~\cite{Oller:1998hw} a somewhat 
low value of $\Gamma(f_0\rightarrow\pi\pi)$=18.2 MeV is obtained. 

It is interesting to note that the analysis of the KLOE collaboration 
on $\phi\rightarrow f_0(980)\gamma\rightarrow 
\pi^+\pi^-\gamma$~\cite{Ambrosino:2005wk} results in the ratio of couplings 
with $R=g^2_{_{f_0K^+K^-}}/g^2_{_{f_0\pi\pi}} = 2.2-2.8$ 
in agreement with our prediction of 
\eq 
R=\frac{g^2_{_{f_0K^+K^-}}}{g^2_{_{f_0\pi\pi}}}=2.28 \,. 
\en

\section{Summary}\label{sec:summary}

We have discussed the electromagnetic $f_0\to\gamma\gamma$ and the strong  
$f_0\to\pi\pi$ decays of the $f_0(980)$ considered as a hadronic $K\bar K$ 
molecule in a phenomenological Lagrangian approach. Our approach is 
manifestly Lorentz and gauge invariant and is based 
on the use of the compositeness condition. We have only one model parameter 
$\Lambda$, which is related to the size of the $K\bar K$ distribution in the 
$f_0$ meson and, therefore, controls finite-size effects. In addition, 
we studied the sensitivity on the detailed value of the binding 
energy $\epsilon$.  

We showed that finite-size effects only moderately influence the 
$f_0 \to \gamma\gamma$ decay properties (coupling constant and decay width), 
when both photons are on-shell (in this point we completely agree with 
the conclusions of ref.~\cite{Hanhart:2007wa}), while these effects become 
essential for a nontrivial virtuality of one of the photons. 
Also, the consideration of the finite size of the $f_0$ meson in the 
molecular $K\bar K$ picture is sufficient to explain the strong 
$f_0\to\pi\pi$ decay properties. More precise experimental information 
on the $f_0\rightarrow\pi\pi$ decay, and more so for the 
$F_{f_0\gamma\gamma^\ast}(Q^2)$ form factor can certainly help in at 
least constraining the molecular $K\bar K$ content of the $f_0(980)$. 
Given the current model approach and the status of experimental data, 
the strong and electromagnetic decay properties of the $f_0(980)$ can 
be fully explained in a molecular $K\bar K$ interpretation of this state.
The success in explaining the decay modes of the $f_0(980)$ does at this 
level not exclude possible alternative explanations, such as $q\bar q$ 
or compact $q^2 \bar q^2$ configurations.

To elaborate further on a possible molecular structure of the $f_0(980)$
in future we plan to include in our analysis the $a_0(980)$ meson 
and its mixing with the $f_0(980)$. Using the approach developed here 
we intend to analyze different strong, radiative and weak production and 
decay processes involving the $f_0(980)$ and $a_0(980)$ mesons.
A full treatment of these observables can possibly
shed more light on the structure of these 
peculiar scalar states.

\vspace*{.5cm} 

\noindent 
{\bf Acknowledgments}

\vspace*{.25cm} 

\noindent 
We thank C.~Hanhart and S.~Scherer for useful discussions.  
This work was supported by the DFG under contracts FA67/31-1 and
GRK683. This research is also part of the EU Integrated
Infrastructure Initiative Hadronphysics project under contract
number RII3-CT-2004-506078 and President grant of Russia
``Scientific Schools''  No. 871.2008.2.

\appendix\section{Mass operator of $f_0(980)$
and coupling constant $g_{f_0K\bar K}$}

The $f_0(980)$ mass operator $\Pi_{f_0}(p^2)$ reads as
\eq
\Pi_{f_0}(p^2) = 2 \int\,\frac{d^4k}{(2\pi)^4i}\,
\widetilde\Phi^2(-k^2)\, S_K(k_+) \,
S_K(k_-) \,, \label{eq: massop}
\en
where $k_\pm = k + p/2$. To simplify the illustration of 
the calculation technique we restrict ourselves to the isospin 
limit. After introducing the Feynman $\alpha$-parametrization with
\eq 
\frac{1}{AB} = \int\limits_0^1d\alpha \frac{1}{(A\alpha + B(1-\alpha))^2}
\en 
and by 
applying the Cauchy theorem for the vertex function 
squared~\cite{Ivanov:1997ug} 
\eq
\widetilde\Phi^2(-k^2)=\oint\frac{dz}{2\pi i} 
\frac{\widetilde\Phi^2(-z)}{z-k^2}
\en 
we get 
\eq 
\Pi_{f_0}(p^2) &=& \frac{1}{8\pi^2} 
\int\limits_0^1 d\alpha \int \, \frac{d^4k}{\pi^2i}\oint\frac{dz}{2\pi i}\frac{\widetilde\Phi(-z)}{z-k^2}\nonumber\\ 
&\times&\frac{1}{(M_K^2 - (k+r)^2 + r^2 - p^2/4)^2} \,, 
\en 
where $r = p (1 - 2 \alpha)/2$. 
Then, using the integral representation 
\eq 
\frac{1}{AB^2} = 2 \int\limits_0^\infty d\beta\,\frac{ \beta}{(A + B\beta)^3} 
\en 
and performing the integration over the loop momentum $k$ we obtain 
\eq 
\Pi(p^2) = \frac{1}{8\pi^2} \int\limits_0^1 d\alpha 
\, \int\limits_0^\infty \frac{d\beta \beta}{(1 + \beta)^2} 
\, \widetilde\Phi^2(\Delta)\,, 
\en 
where 
\eq
\Delta = \beta \biggl( M_K^2 - p^2 
\frac{1 + 4\alpha \beta (1 - \alpha)}{4 (1 + \beta)} 
\biggr)
\,.
\en 
Finally, by evaluating the derivative of the mass operator with respect to 
$p^2$ for the on-mass-shell value of $p^2 = M_{f_0}^2$ we calculate the 
coupling constant $g_{f_0 K \bar K}$ by use of the compositeness 
condition~(\ref{Zf0}) with: 
\eq\label{gf0KK} 
\frac{1}{g_{f_0 K \bar K}^2} &=& \frac{1}{32\pi^2 \Lambda^2} 
\int\limits_0^1 d\alpha 
\, \int\limits_0^\infty \frac{d\beta \beta^2}{(1 + \beta)^3} 
\, ( 1 + 4 \alpha \beta (1 - \alpha) )\nonumber\\ 
&\times&( 1 + 4 \alpha \beta (1 - \alpha) )\biggl( - \frac{d\widetilde\Phi^2(\Delta)}{d\Delta} \biggr)
\bigg|_{p^2 = M_{f_0}^2} \,.
\en 
In the local limit, that is $\Lambda\rightarrow\infty$, 
eq.~(\ref{gf0KK}) becomes 
\eq 
\frac{1}{g_{f_0K\bar K}^2} = \frac{R_{f_0}}{8\pi^2 M_{f_0}^2}
\en 
with 
\eq 
R_{f_0} = \frac{\arcsin\xi}{\xi\sqrt{1-\xi^2}} - 1 \,.  
\en 

\section{Feynman rules for the nonlocal vertices} 

Restoration of electromagnetic gauge invariance in the strong 
$f_0 K \bar K$ interaction Lagrangian modifies the coupling 
to charged kaons as 
\eq\label{L_str_em2}
{\cal L}(x) &=& 
g_{f_0K\bar K} f_0(x)\!\! \int dy \,\Phi(y) \, 
e^{-ieI(x_+,x_-)} \nonumber\\
&\times& K^+(x_+) \, K^-(x_-) \,, 
\en 
where $x_\pm = x \pm y/2$. \newline
After expansion of the gauge exponentials up to second order in the 
electromagnetic field we generate the nonlocal vertices containing 
a single photon [fig.~2(a)] and two photon lines [fig.~2(b)]. 
In our procedure we follow ref.~\cite{Faessler:2003yf} where we obtain 
for the single-photon vertex 
\eq 
e\left(k+\frac {q_1}{4}\right)^\mu\int\limits_0^1dt
\,\widetilde\Phi^\prime(x(0,q_1))+(q_1\leftrightarrow q_2,\;
\mu\leftrightarrow\nu)\,, \label{eq: fr1}
\en
and for the two-photon vertex 
\begin{eqnarray}
&-&e^2\frac{g^{\mu\nu}}{4}\int\limits_0^1dt
\left( \, \widetilde\Phi^\prime(-x(0,\,p))
+\widetilde\Phi^\prime(-x(0,\,q)) \, \right) \nonumber\\
&+&\int\limits_0^1dt\,t\int\limits_0^1 dl
\left( \, \widetilde\Phi^{\prime\prime}(-x(q_1,\,q_2)) 
\left(k+q_{21}^+\right)^\mu
\left(k+\frac{q_4}{4}\right)^\nu\right.\nonumber\\
&+&\left.\widetilde\Phi^{\prime\prime}(-x(-q_1,\,q_2))
\left(k+q_{21}^-\right)^\mu
\left(k+\frac{q_2}{4}\right)^\nu \, \right)\nonumber\\
&+&(q_1\leftrightarrow q_2,\;\mu\leftrightarrow\nu)\,, \label{eq: fr2}
\end{eqnarray} 
where 
\eq
x(q_1,q_2) \, &=& \,  k^2 + k t (l\,q_1+q_2)
            +\frac{t}{4}(l\,q_1^2+2\,l\,q_1q_2+q_2^2)\nonumber\\
q_{ij}^\pm &=& \frac{q_i}{2} \pm \frac{q_j}{4}\nonumber\\
q \,& =& \, q_2 - q_1 \,. \label{eq: x1}
\en

\section{Form factors and gauge invariance of local diagrams} 

We divide the hadron loop integral
\eq 
I^{L \, \mu\nu}_{\triangle}(q_1,q_2) 
&=&\int\,\frac{d^4 k}{\pi^2i}\,(2k+q_1)^\mu\,(2k-q_2)^\nu\nonumber\\
&\times&\,S(k+q_1)
\,S(k-q_2)\,S(k)\,,            \label{eq: t1}
\en
represented by the diagram in fig.~4(a), into a gauge invariant part and 
the remainder term
\eq 
&&I^{L \, \mu\nu}_{\triangle_\perp}(q_1,q_2) + 
\delta I^{L \, \mu\nu}_{\triangle}(q_1,q_2)\!\nonumber\\
 &&= 
\!\int\,\frac{d^4 k}{\pi^2i}(2k+q_1)_{\perp\,q_1}^\mu 
\,(2k-q_2)_{\perp\,q_2}^\nu \nonumber\\
&&\times S(k+q_1)S(k-q_2)S(k)\nonumber\\
&&+\left(\frac{c^{\mu\nu}}{q_1^2q_2^2}-g^{\mu\nu}\right) 
\,\int\frac{d^4k}{\pi^2i}S(k+q_1)\,S(k-q_2)\;\label{eq: t2}.
\en 
Here and in the following (Appendices C and D) we only deal with 
the propagator of charged kaons. We therefore drop in the following 
the subscript $K^+$ in the symbol of the propagator.

The second diagram of fig.~4(b) gives
\eq 
I^{L \, \mu\nu}_{\circ}(q_1,q_2)
=  g^{\mu\nu} \, \int\frac{d^4k}{\pi^2i}S(k+q_1)\,S(k-q_2) \,, 
\label{eq: t3}
\en
where (\ref{eq: t3}) and the last term in (\ref{eq: t2}) cancel 
each other. Therefore, we can easily derive the form factors
\eq 
&&I^{L \, \mu\nu}_{\triangle_\perp}(q_1,q_2) \nonumber\\
&&= \int\,\frac{d^4 k}{\pi^2i}\,(2k+q_1)_{\perp\,q_1}^\mu 
\,(2k-q_2)_{\perp\,q_2}^\nu\,\nonumber\\
&&\times S(k+q_1)\,S(k-q_2)\,S(k)\nonumber\\
&&+\frac{c^{\mu\nu}}{q_1^2q_2^2}  \, 
\int\,\frac{d^4 k}{\pi^2i}\,S(k+q_1)S(k-q_2) \nonumber\\
&&=F_{\triangle,\,L}(p^2,q_1^2,q_2^2) \, b^{\mu\nu} + 
G_{\triangle,\,L}(p^2,q_1^2,q_2^2) \, c^{\mu\nu}\,.
\en 
By introducing the Feynman $\alpha$-parameters and using dimensional 
regularization we obtain
\eq 
F^L_{\triangle}(p^2,q_1^2,q_2^2) &=&
\int\limits_0^1d^3\alpha\,\delta(1-\sum\limits_{i=1}^3\alpha_i)
\,\frac{4\alpha_1\alpha_2}{D}\,, \label{eq: floc}\\
&&\nonumber\\
G^L_{\triangle}(p^2,q_1^2,q_2^2) &=& \frac{1}{q_1^2q_2^2} \,  
\int\limits_0^1d^3\alpha\,\delta(1-\sum\limits_{i=1}^3\alpha_i)
\nonumber\\
&\times&\left( 2 \ln\frac{D}{D_0}-\frac{4\alpha_1\alpha_2}{D}
q_1q_2 \, \right) \,,  
\label{eq: gloc}
\en 
where 
\eq 
D&=&M^2_{K}-p^2\alpha_1\alpha_2-q_1^2\alpha_1\alpha_3 
-q_2^2\alpha_2\alpha_3 \,, \nonumber\\
D_0&=&M^2_{K}-p^2\alpha_1\alpha_2\,.
\en 

\section{Form factors and gauge-invariance of 
the nonlocal diagrams} 

\subsection{Triangle diagram}
In Appendix C we saw that there are two diagrams which belong to the 
leading-order term illustrated in fig.~4. In complete analogy to the local 
case we write down the Feynman integral and split it into a gauge invariant 
and a remainder term. The first diagram of fig.~4(a) gives  
\eq 
&&I^{\mu\nu}_{\triangle}(q_1,q_2)\nonumber\\
&&=\int\frac{d^4k}{\pi^2i}\,\widetilde\Phi(-k^2)\left(2k+q_2\right)^\mu\left(2k-q_1\right)^\nu\nonumber\\
&&\;\times S\left(k+\frac {p}{2}\right)\,S\left(k-\frac {p}{2}\right)
\,S\left(k+\frac {q}{2}\right)\nonumber\\
&&=\int\frac{d^4k}{\pi^2i}\,\widetilde\Phi(-k^2)\,
\left(2k+q_2\right)^\mu_{\perp \,q_1}\left(2k-q_1\right)^\nu_{\perp \, q_2}\nonumber\\
&&\;\times S\left(k+\frac {p}{2}\right)\,S\left(k-\frac {p}{2}\right)
S\left(k+\frac {q}{2}\right)\nonumber\\
&&\;+\int\frac{d^4k}{\pi^2i}
\widetilde\Phi\left(-\left(k-\frac{q_2}{2}\right)^2\right)
\left(\frac{2k^\mu_{\perp q_1}q^\nu_2}{q_2^2}
+\frac{q_1^\mu q_2^\nu}{q_1^2q_2^2}kq_1\right) \nonumber\\
&&\;\times S\left(k+\frac{q_1}{2}\right)S\left(k-\frac{q_1}{2}\right)\nonumber\\ 
&&\;-\int\frac{d^4k}{\pi^2i}\widetilde\Phi\left(-k^2\right)
\left(g^{\mu\nu}-\frac{c^{\mu\nu}}{q_1^2q_2^2}\right)
S\left(k+\frac{p}{2}\right)S\left(k-\frac{p}{2}\right)\nonumber\\
&&-\int\frac{d^4k}{\pi^2i}
\widetilde\Phi\left(-\left(k+\frac{q_1}{2}\right)^2\right)
\left(\frac{2q_1^\mu k^\nu_{\perp q_2}}{q_1^2}
+\frac{q_1^\mu q_2^\nu}{q_1^2q_2^2}kq_2\right) \nonumber\\
&&\;\times S\left(k+\frac{q_2}{2}\right)S\left(k-\frac{q_2}{2}\right)\,.
\en 
The second diagram of fig.~4(b) contributes
\eq
I^{\mu\nu}_{\circ}(q_1,q_2) = g^{\mu\nu} \int\frac{d^4k}{\pi^2i}
\,\widetilde\Phi(-k^2)\,S\left(k-\frac{p}{2}\right)
S\left(k+\frac{p}{2}\right) \,.\nonumber
\en
By writing down the invariant part of the integrals in terms of the 
tensor structures $b^{\mu\nu}$ and $c^{\mu\nu}$, one obtains the form factors
\eq 
&&F_{\triangle_\perp}(p^2,q_1^2,q_2^2)\nonumber\\
&&= - \int\limits_0^1d^3\alpha\, 
\delta\left(1-\sum\limits_{i=1}^3\alpha_i\right)\int\limits_0^1 
\frac{dt t^2}{(1-t)^2} \nonumber\\
&&\times\widetilde\Phi^\prime(\Delta_1) \, 
( 1 - 2 t \alpha_3 + t^2 \alpha_{132} \alpha_{231} ) \,, 
\label{eq: ftri}\\
&&G_{\triangle_\perp}(p^2,q_1^2,q_2^2)\nonumber\\
&&= 
\displaystyle\frac{q_1q_2}{q_1^2q_2^2} \, 
\int\limits_0^1d^3\alpha\,\delta 
\left(1-\sum\limits_{i=1}^3\alpha_i\right)\int\limits_0^1\frac{dt t^2}{(1-t)^2}   \nonumber\\
&&\times\, 
\biggl( - 2 \widetilde\Phi(\Delta_1) 
+\widetilde\Phi^\prime(\Delta_1) \, 
\frac{1 - 2 t \alpha_3 + t^2 \alpha_{132} \alpha_{231}}{1 - t} 
\biggr)\,,    \label{eq: gtri}
\en 
where 
\eq 
\Delta_1 &=& \frac{t}{1-t} 
\biggl( M_K^2 - \frac{p^2}{4} ( \alpha_{123} + t \alpha_{132} \alpha_{231} )\nonumber\\  
&-& \frac{q_1^2 \alpha_3}{2} ( 1 - t \alpha_{231}) 
- \frac{q_2^2 \alpha_3}{2} ( 1 - t \alpha_{132}) \biggr)
\en 
where $\alpha_{ijk} = \alpha_i + \alpha_j - \alpha_k$. 

The remainder term reads as 
\eq 
&&\delta I^{\mu\nu}_{\triangle}(q_1,q_2)\nonumber\\
&&=\int\frac{d^4k}{\pi^2i}\,\widetilde\Phi(-k^2) \nonumber\\ 
&&\times\Biggl( \, \frac{q_1^\mu q_2^\nu}{q_1^2q_2^2}  
\, \biggl( \, S\left(k+\frac q2\right) - 
S\left(k-\frac p2\right)\biggr) 
\nonumber\\
&&-\frac{q_1^\mu(2k-q_1)^\nu}{q_1^2} \, 
S\left(k+\frac q2\right) \, S\left(k-\frac p2\right) \nonumber\\
&&+ \frac{(2k+q_2)^\mu q_2^\nu}{q_2^2} \, S\left(k+\frac q2\right)
\, S\left(k+\frac p2\right)\Biggr)\,.\label{eq: remtri}
\en 

\subsection{Bubble diagram}
By writing the Feynman rule for nonlocal vertices, derived in Appendix B, 
and separating the perpendicular term, we have
\eq 
&&I^{\mu\nu}_{\rm bub}(q_1,q_2)\nonumber\\
&&= - 2 \,\int\!\frac{d^4 k}{\pi^2 i}\int\limits_0^1\!d t\,
\widetilde\Phi^\prime(- x(0,q_1)) 
\left(k+\frac{q_1}{4}\right)^\mu k^\nu \nonumber\\
&&\times S\left(k+\frac {q_2}{2}\right)\,S\left(k-\frac {q_2}{2}\right)\nonumber\\
&&+(q_1\,\leftrightarrow\,q_2,\;\mu\,\leftrightarrow\,\nu)\nonumber\\
&&=- 2 \, \int\!\frac{d^4 k}{\pi^2 i}\int\limits_0^1\!d t\,
\widetilde\Phi^\prime(- x(0,q_1))\, 
k^\mu_{\perp \, q_1}k^\nu_{\perp \, q_2}\nonumber\\
&&\times S\left(k+\frac {q_2}{2}\right)
\,S\left(k-\frac {q_2}{2}\right)\nonumber\\
&&+ 2 \, \int\,\frac{d^4 k}{\pi^2 i}\,
\frac{\widetilde\Phi(-(k+\frac{q_1}{2})^2)
-\widetilde \Phi(-k^2)}{kq_1+\frac{q_1^2}{4}}\nonumber\\
&&\times S\left(k+\frac{q_2}{2}\right)S\left(k-\frac{q_2}{2}\right) \nonumber\\ 
&&\times\left( \, \left(k+\frac{q_1}{4}\right)^\mu q_2^\nu \,
\frac{kq_2}{q_2^2}+q_1^\mu k^\nu\,
\frac{\left(k+\frac{q_1}{4}\right)q_1}{q_1^2}\right.\nonumber\\
&&-\left.\, q_1^\mu q_2^\nu
\frac{\left(k+\frac{q_1}{4}\right)q_1}{q_1^2q_2^2}kq_2\right)\nonumber\\
&&+(q_1\,\leftrightarrow\,q_2,\;\mu\,\leftrightarrow\,\nu)
\en 
\eq 
&&F_{{\rm bub}_\perp}(p^2,q_1^2,q_2^2)\nonumber\\
&&=\frac{1}{2} \int\limits_0^1dt t 
\int\limits_0^1d\alpha (2\alpha-1) \int\limits_0^\infty 
\frac{d\beta \beta^2}{(1+\beta)^4} \, 
\widetilde\Phi^\prime(\Delta_2)\nonumber\\
&&+(q_1\,\leftrightarrow\,q_2,\;
\mu\,\leftrightarrow\,\nu) \,, \label{eq: fbub}\\
&&G_{{\rm bub}_\perp}(p^2,q_1^2,q_2^2)=\frac{1}{q_1^2q_2^2} \, 
\int\limits_0^1dt \int\limits_0^1d\alpha\int\limits_0^\infty \, 
\frac{d\beta \beta}{(1+\beta)^4} \nonumber\\
&&\times\left\{ \,\widetilde\Phi(\Delta_2)
  +\frac{1}{2} \, t \, \beta \, (2\alpha-1) \, 
\widetilde\Phi^\prime(\Delta_2) \, q_1q_2\right\} 
\nonumber\\
&&+\,(q_1\,\leftrightarrow\,q_2,\;\mu\,\leftrightarrow\,\nu)\,,\label{eq: gbub}
\en 
where
\eq 
\Delta_2 &=& M_K^2 \beta 
+ p^2 \, \frac{t \beta (1 - 2\alpha)}{4 (1 + \beta)}  
- q_1^2 \, 
\frac{t (1 - t + 2 \beta (1 - \alpha))}{4 (1 + \beta)} \nonumber\\  
&-& q_2^2  \, \frac{ \beta (1 + t (1 - 2 \alpha) 
+ 4 \alpha \beta (1 - \alpha)))}{4 (1 + \beta)} \,. 
\en 
The remainder term is given by
\eq 
&&\delta I^{\mu\nu}_{\rm bub}(q_1,q_2)\nonumber\\
&&= \frac{q_1^\mu}{q_1^2} 
\int\frac{d^4k}{\pi^2i}\,\widetilde\Phi(-k^2) \, 
\Big( \, \frac{q_2^\nu}{2 q_2^2}
\left(S\left(k-\frac p2\right)-S\left(k+\frac q2\right)\right)\nonumber\\
&&+ 2 (k-\frac{q_1}{2})^\nu
S\left(k+\frac q2\right)S\left(k-\frac p2\right) \, \Big)\nonumber\\
&&+ \frac{q_1^\mu}{q_1^2} \int\frac{d^4k}{\pi^2i}\,S(k) \, 
\Biggl( \widetilde\Phi\left(-\left(k+\frac{q}{2}\right)^2\right) 
\,\left( \frac{(k - q_{12}^-)^\nu}{(k - q_{12}^-) q_2}\right.
\nonumber\\
&&\left.-\frac{q_2^\nu}{2q_2^2}\right)-\widetilde\Phi\left(-\left(k+\frac{p}{2}\right)^2\right)
\,\left( \frac{(k + q_{12}^+)^\nu}{(k + q_{12}^+) q_2} 
- \frac{q_2^\nu}{2q_2^2} \right)\nonumber\\[4mm] 
&&+\widetilde\Phi\left(-\left(k+\frac{q_1}{2}\right)^2\right)
\,\left( \frac{(k + q_{12}^+)^\nu}{(k + q_{12}^+) q_2} 
- \frac{(k + q_{12}^-)^\nu}{(k + q_{12}^-) q_2} \right) \, \Biggr) \nonumber\\
&+& \, (q_1\,\leftrightarrow\,q_2,\;\mu\,\leftrightarrow\,\nu)\,. 
\label{eq: rembub}
\en 

\subsection{Tadpole diagram}

We apply the same procedure as for the bubble diagram and write down 
the Feynman integral 
\eq 
I^{\mu\nu}_{\rm tad}(q_1,q_2)=\int\frac{d^4k}{\pi^2i}\,S(k)\Lambda^{\mu\nu}\,,
\en 
where
\eq 
\Lambda^{\mu\nu}&=&-\frac{g^{\mu\nu}}{4}\int\limits_0^1 dt \, 
\biggl( \, \widetilde\Phi^\prime(-x(0,\,p))\nonumber\\
&+&
\widetilde\Phi^\prime(-x(0,\,q)) \, \biggr) \nonumber\\ 
&+& \int\limits_0^1dt\,t\int\limits_0^1 dl \, \biggl(
\widetilde\Phi^{\prime\prime}(-x(q_1,\,q_2)) 
\, \left(k + q_{21}^+\right)^\mu \nonumber\\
&+& \widetilde\Phi^{\prime\prime}(-x(-q_1,\,q_2))
\left(k + q_{21}^-\right)^\mu 
\biggr) \left(k+\frac{q_2}{4}\right)^\nu  \nonumber\\
&+&(q_1\,\leftrightarrow\,q_2,\;\mu\,\leftrightarrow\,\nu) \nonumber\\ 
&=&I^{\mu\nu}_{\rm tad_\perp}+\delta I^{\mu\nu}_{\rm tad}\,.
\en 
By using 
\eq 
g^{\mu\nu}=g^{\mu\nu}_\perp+\left(\frac{q_1^\mu q_1^\nu}{q_1^2}
+\frac{q_2^\mu q_2^\nu}{q_2^2}-\frac{q_1^\mu 
q_2^\nu \, q_1q_2}{q_1^2q_2^2}\right)
\en
we can separate the gauge invariant part
\eq 
&&I^{\mu\nu}_{\rm tad_\perp}(q_1,q_2)\nonumber\\
&&=\int\frac{d^4k}{\pi^2i}\,S(k) \, \Biggl( \left(
k+\frac{q_2}{2}\right)^\mu_{\perp q_1}k^\nu_{\perp q_2} 
\nonumber\\
&&\times \int\limits_0^1dt\,t\int\limits_0^1dl \,
\left(\widetilde\Phi^{\prime\prime}(-x(q_1,\,q_2))+
  \widetilde\Phi^{\prime\prime}(-x(-q_1,\,q_2)) \right)\nonumber\\
&&-\frac{c^{\mu\nu}}{4q_1^2q_2^2}
\int\limits_0^1dt\,\left( \widetilde\Phi^\prime(-x(0,\,p))
+\widetilde\Phi^\prime(-x(0,\,q))\right) \Biggr)  
\nonumber\\
&&+ \, (q_1\,\leftrightarrow\,q_2,\;\mu\,\leftrightarrow\,\nu)\,.
\en 
from which we derive the form factors
\eq 
&&F_{\rm tad_\perp}(p^2,q_1^2,q_2^2)\nonumber\\
&&= \frac{1}{4} 
\int\limits_0^1 dt t^2\int\limits_0^1dl l 
\int\limits_0^\infty \frac{d\beta}{(1+\beta)^3}\nonumber\\  
&&\times\left(\widetilde\Phi^\prime(\Delta_3)
-\widetilde\Phi^\prime(\Delta_4)\right) \, 
\left(\frac{t}{1+\beta}-1\right)    \nonumber\\
&&+(q_1\,\leftrightarrow\,q_2,\;\mu\,\leftrightarrow\,\nu) \,,       
\label{eq: ftad}\\
&&\nonumber\\
&&G_{\rm tad_\perp}(p^2,q_1^2,q_2^2)\nonumber\\
&&=\frac{1}{4q_1^2q_2^2} 
\int\limits_0^1 dt t\int\limits_0^1dl\, 
\int\limits_0^\infty \frac{d\beta}{(1+\beta)^3} 
\Biggl( \widetilde\Phi(\Delta_3) - \widetilde\Phi(\Delta_4) \nonumber\\
&&- q_1q_2 \, \left(\widetilde\Phi^\prime(\Delta_3) 
- \widetilde\Phi^\prime(\Delta_4)\right) \, 
t^2 l \, \left(\frac{t}{1+\beta}-1\right) \Biggr) \nonumber\\
&&+ \frac{1}{4q_1^2q_2^2} 
\int\limits_0^1dt\,\int\limits_0^\infty \frac{d\beta}{(1+\beta)^2}
\left(\widetilde\Phi(-\Delta_5)+\widetilde\Phi(-\Delta_6)\right)\nonumber\\
&&+(q_1\,\leftrightarrow\,q_2,\;\mu\,\leftrightarrow\,\nu)\,,   
\label{eq: gtad}
\en 
where
\eq 
\Delta_3&=& M_K^2 \, \beta 
+ p^2 \, \frac{tl}{4} \left(\frac{t}{1+\beta}- 1\right)\nonumber\\
&-& q_1^2 \, \frac{t^2l(1 - l)}{4 (1+\beta)} 
- q_2^2 \, \frac{t(1-l)}{4}\left( 1 - \frac{t}{1+\beta}\right) 
\,, \nonumber\\
\Delta_4&=& M_K^2 \, \beta 
+ p^2 \, \frac{tl}{4} 
\left( 1 -\frac{t}{1+\beta} \right) \nonumber\\
&-& q_1^2 \, \frac{tl}{4} \left( 2 - \frac{(1+l)t}{1+\beta}\right)
- q_2^2 \, \frac{t(1+l)}{4}\left(1 - \frac{t}{1+\beta}\right)\,,  
\nonumber\\
\Delta_5&=& M_K^2 \, \beta 
+ p^2 \, \frac{t}{4} \left(\frac{t}{1+\beta} - 1 \right)\,, 
\nonumber\\ 
\Delta_6&=& M_K^2 \, \beta + (2q_1^2+2q_2^2-p^2) \, \frac{t}{4} 
\left(\frac{t}{1+\beta} - 1\right) \,. 
\en
For the remainder term we obtain
\eq 
\delta I^{\mu\nu}_{tad}(q_1,q_2)=\int\frac{d^4k}{\pi^2i}S(k)\,R^{\mu\nu}\,,
\en 
where
\eq 
R^{\mu\nu}&=&\;\widetilde\Phi\left(-\left(k+\frac p2\right)^2\right) \, 
\left(\frac{q_1^\mu (k + q_{12}^+)^\nu}{q_1^2 (k + q_{12}^+)q_2} 
- \frac{q_1^\mu q_2^\nu}{2q_1^2q_2^2} \right) \nonumber\\
&+&\widetilde\Phi\left(-\left(k+\frac q2\right)^2\right)
\left(-\frac{q_1^\mu (k - q_{12}^-)^\nu}{q_1^2 (k - q_{12}^-)q_2}  
+ \frac{q_1^\mu q_2^\nu}{2q_1^2q_2^2}\right) \nonumber\\[3mm] 
&+&\widetilde\Phi\left(-\left(k+ \frac{q_1}{2}\right)^2\right)
\frac{q_1^\mu}{q_1^2}\left( \frac{(k + q_{12}^-)^\nu}{(k + q_{12}^-)q_2} 
-\frac{(k + q_{12}^+)^\nu}{(k + q_{12}^+)q_2} \right) \nonumber\\ 
&+&(q_1\,\leftrightarrow\,q_2,\;\mu\,\leftrightarrow\,\nu)\,. 
\label{eq: remtad}
\en 
The sum of all remainder terms (\ref{eq: remtri}), (\ref{eq: rembub}) 
and (\ref{eq: remtad}) vanishes identically, which in turn proofs the manifest 
gauge invariance of the model.

\newpage 


\begin{table*}
\caption{Electromagnetic decay width $f_0(980)\to\gamma\gamma$: 
comparison with data and other approaches (quarkonia models ($q\bar q$), 
four-quark models ($q^2\bar q^2$) and hadronic approaches).}   
\begin{center}
\def\arraystretch{2} 
\begin{tabular}{|c|c|c|c|c|}
\hline 
Approach & Data~\cite{Mori:2006jj}
         & Data~\cite{Yao:2006px} 
         & Data~\cite{Marsiske:1990hx}
         & Data~\cite{Boyer:1990vu} \\
\hline 
$\Gamma(f_0\to\gamma\gamma)$, keV 
     & $\;\;0.205^{+0.095\,+0.147}_{-0.083\,-0.117}\;\;$
                                  & $\;\;0.29^{+0.07}_{-0.09}\;$ 
                                  & $\;\;0.31\pm0.14\pm0.09\;\;$
                                  & $\;\;0.29\pm0.07\pm0.12\;\;$ \\
\hline
\end{tabular}
\end{center}
\begin{center}
\def\arraystretch{2}
\hspace*{0.05cm}
\begin{tabular}{|c|c|c|c|c|c|c|c|c|}
\hline 
Approach & Ref.~\cite{Oller:1997yg}
         & Ref.~\cite{Hanhart:2007wa}
         & Ref.~\cite{Efimov:1993ei}
         & Ref.~\cite{Anisovich:2001zp} 
         & Ref.~\cite{Scadron:2003yg}
         & Ref.~\cite{Schumacher:2006cy} 
         & Ref.~\cite{Achasov:1981kh} 
         & Our \\
         &(hadronic)
         &(hadronic)
         &$(q\bar q)$
         &$(q\bar q)$
         &$(q\bar q)$
         &$(q\bar q)$
         &$(q^2\bar q^2)$
         &result\\
\hline
$\Gamma(f_0\to\gamma\gamma)$, keV & 0.20
                                  & $0.22 \pm 0.07$ 
                                  & 0.24 
                                  & $0.28^{+0.09}_{-0.13}$
                                  & 0.31 
                                  & 0.33 
                                  & 0.27
                                  & 0.29 (local)  \\
&&&&&&&&0.25 (nonlocal)\\
\hline
\end{tabular}
\label{tab:em}
\end{center}

\vspace*{2cm}

\caption{Strong decay width $f_0 \to \pi\pi$:  
comparison with data and other approaches.}  
\begin{center}
\def\arraystretch{2}
\begin{tabular}{|c|c|c|c|c|c|c|c|c|c|c|}
\hline 
Approach & Data~\cite{Mori:2006jj} 
         & Data~\cite{Yao:2006px}
         & Ref.~\cite{Anisovich:2001ay}
         & Ref.~\cite{Oller:1998hw}
         & Ref.~\cite{Efimov:1993ei}
        \\
        &(Belle)&(PDG)&(analysis)&(hadronic)&($q\bar q$)\\
\hline 
$\Gamma(f_0\to\pi\pi)$, MeV & $51.3^{+20.8\,+13.2}_{-17.7\,-3.8}$ 
                            & 40 - 100 
                            & $64\pm 8$
                            & 18.2 
                            & 20 
                            \\
\hline
\end{tabular}\\
\vspace*{0.2cm}
\begin{tabular}{|c|c|c|c|c|c|c|c|c|c|c|}
\hline 
Approach & Ref.~\cite{Volkov:2000vy}
          &Ref.~\cite{Anisovich:2002ij}
         & Ref.~\cite{Scadron:2003yg}
         & Ref.~\cite{Celenza:2000uk}
         & Our \\
         &$(q\bar q)$&$(q\bar q)$&$(q\bar q)$&$(q\bar q)$&result\\
\hline 
$\Gamma(f_0\to\pi\pi)$, MeV & 28 
                            &52-58 
                            & 53 
                            & 56
                            & 69 \\
\hline
\end{tabular}
\label{tab:strong}
\end{center}
\end{table*}

\newpage 
\newpage


\begin{figure*}
\centering{\
\epsfig{figure=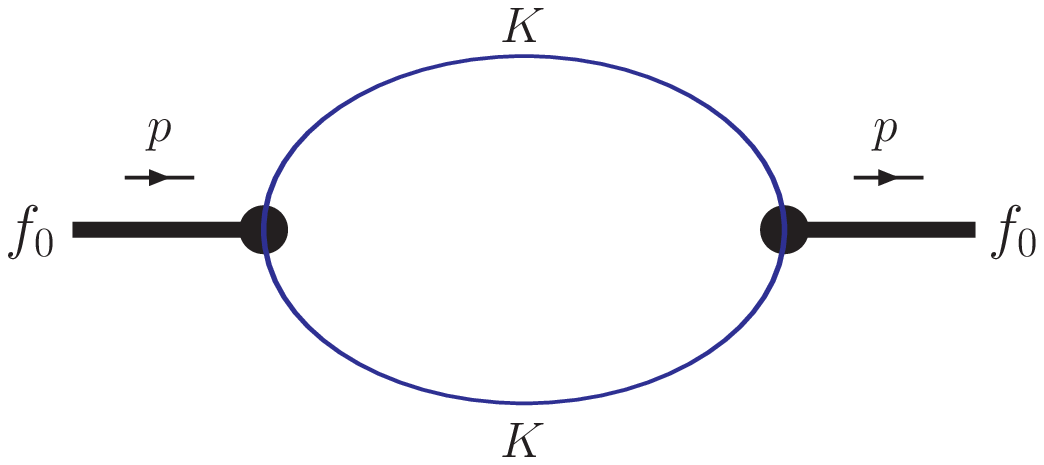,scale=.7}}
\caption{Mass operator of the $f_0$ meson.} 
\label{fig:massop}

\vspace*{1cm}

\centering{\
\epsfig{figure=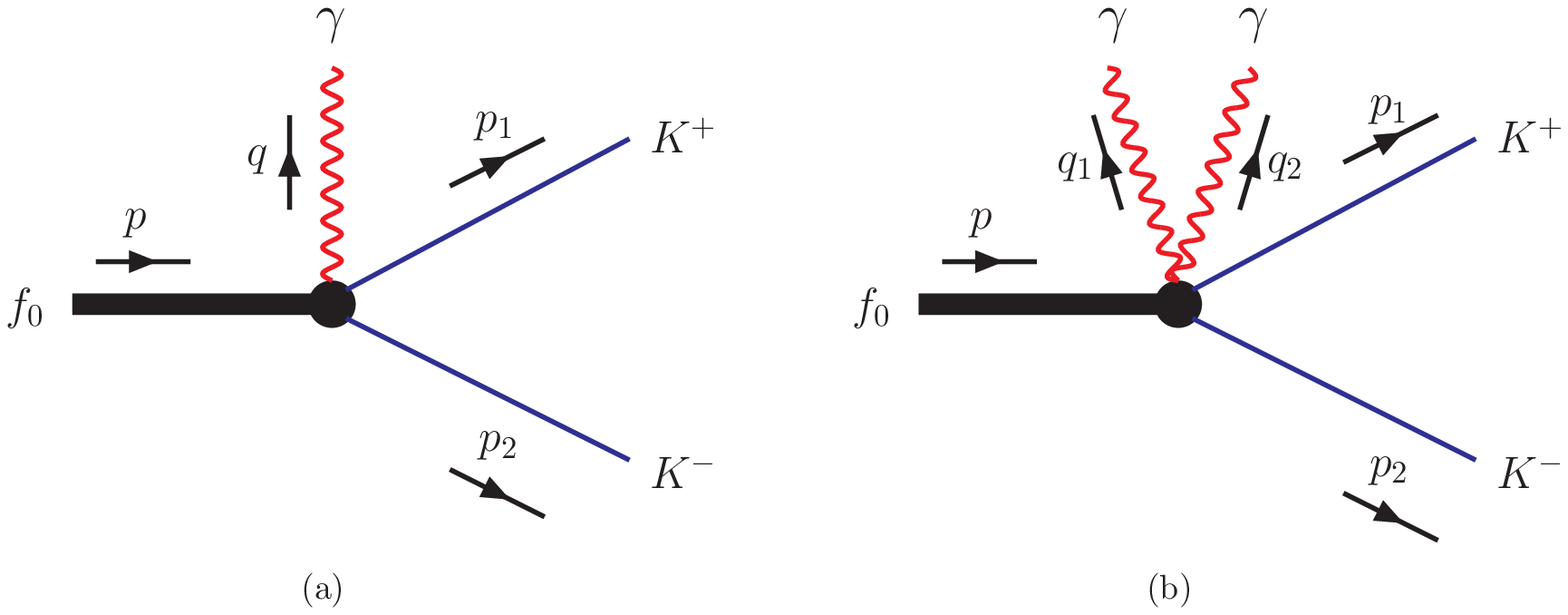,scale=.6}}
\caption{Electromagnetic vertices generated by the 
restoration of gauge invariance in the nonlocal case.}
\label{fig:strong}

\vspace*{1cm}

\centering{\
\epsfig{figure=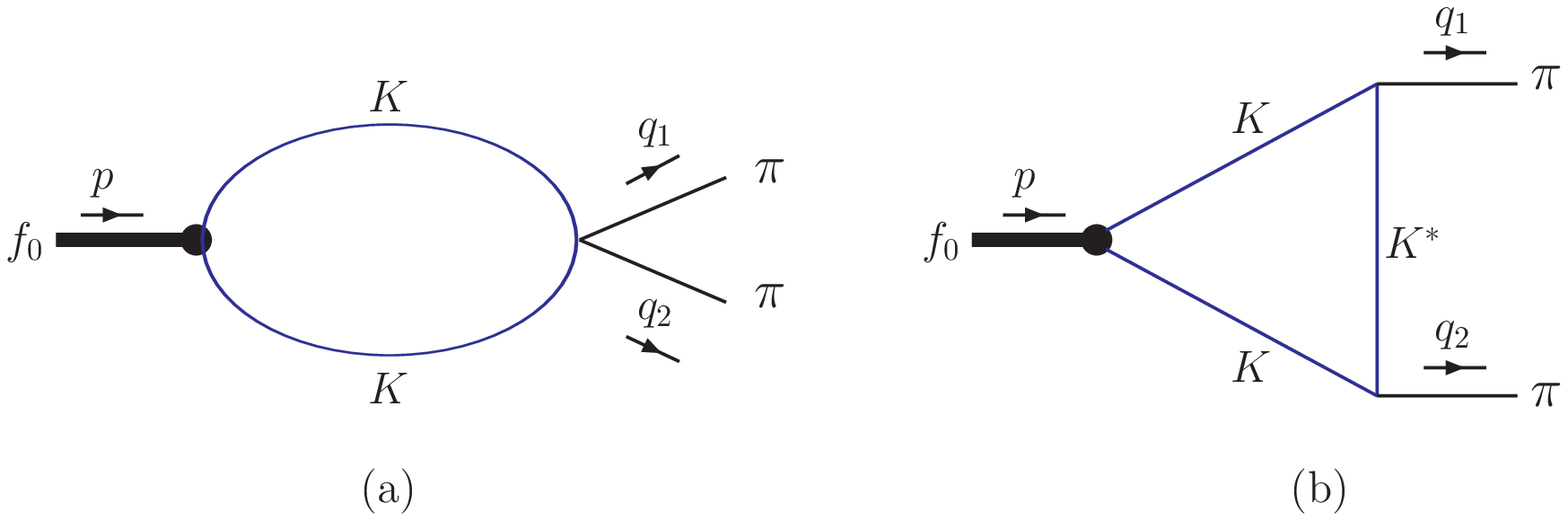,scale=.8}}
\caption{Diagrams contributing to the strong $f_0 \to \pi\pi$ decay.}
\label{fig:em}
\end{figure*}

\begin{figure*} 
\centering{\
\epsfig{figure=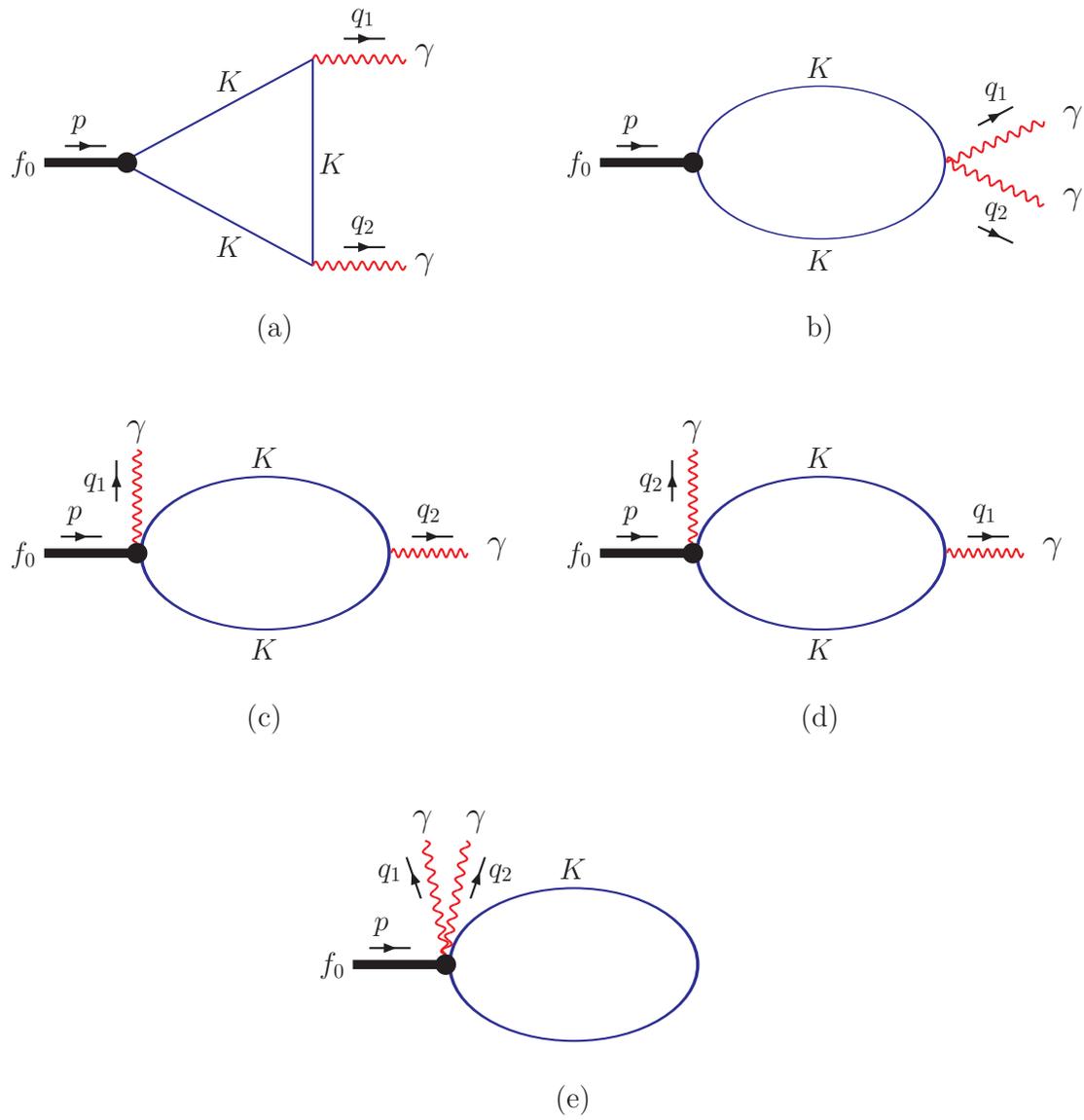,scale=.8}}
\caption{Diagrams contributing to the electromagnetic 
$f_0 \to \gamma\gamma$ decay.}
\label{fig:vertex}
\end{figure*}

\newpage

\begin{figure*}
\centering{\
\epsfig{figure=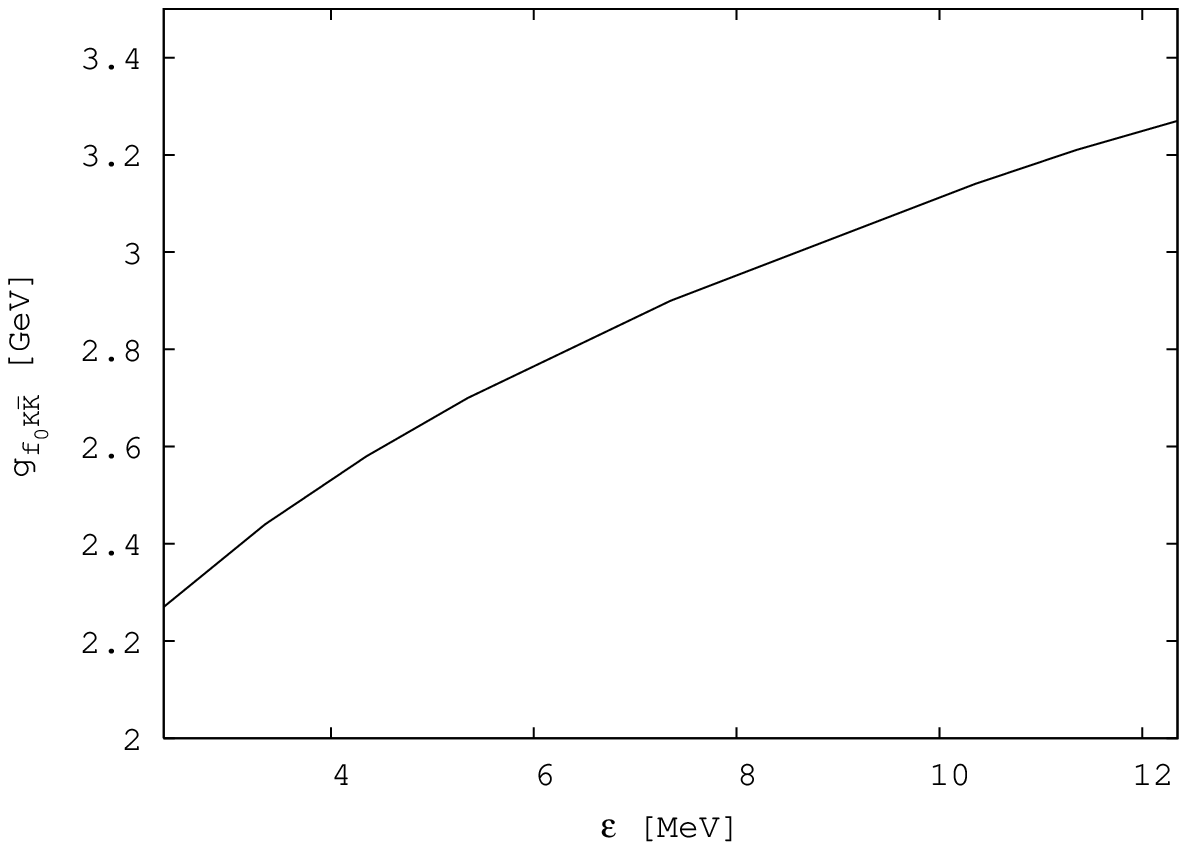,height=9cm}}
\caption{Coupling constant $g_{f_0 K \bar K}$ in the local case 
in dependence on the binding energy $\epsilon$.}

\vspace*{1.3cm} 

\centering{\
\epsfig{figure=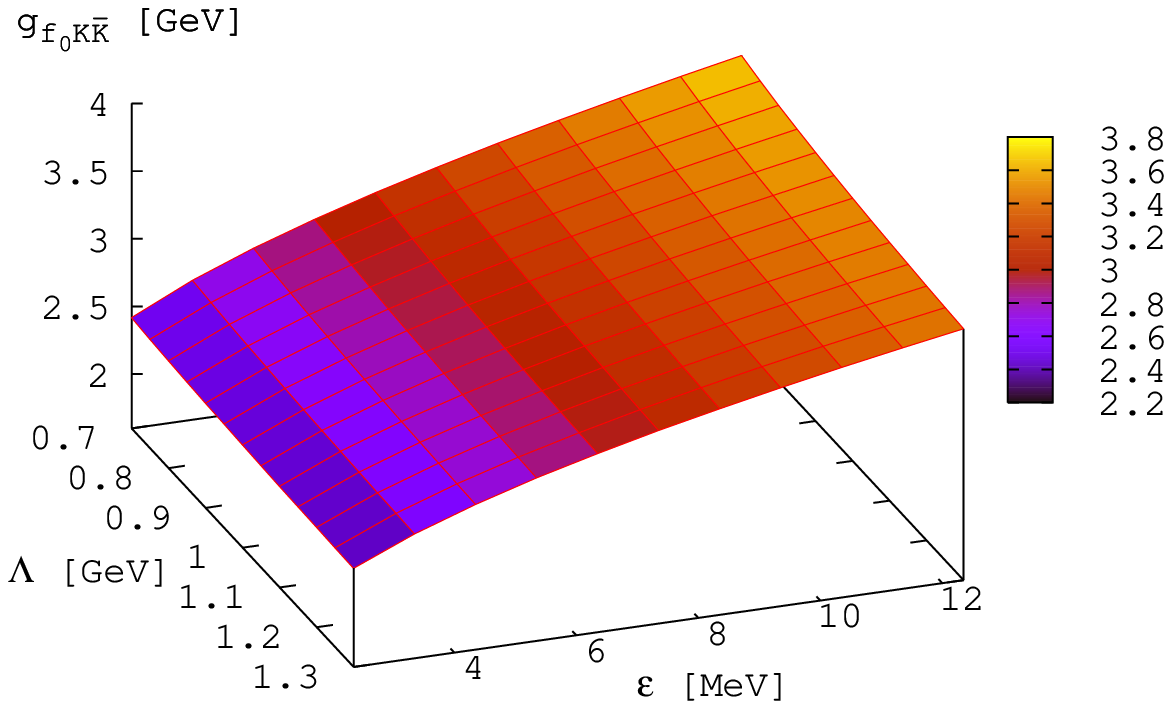,height=10cm}}
\caption{Coupling constant $g_{f_0 K \bar K}$ in the nonlocal case \
in dependence on the binding energy $\epsilon$ and the cut-off $\Lambda$.}
\end{figure*}

\newpage

\begin{figure*}
\centering{\
\epsfig{figure=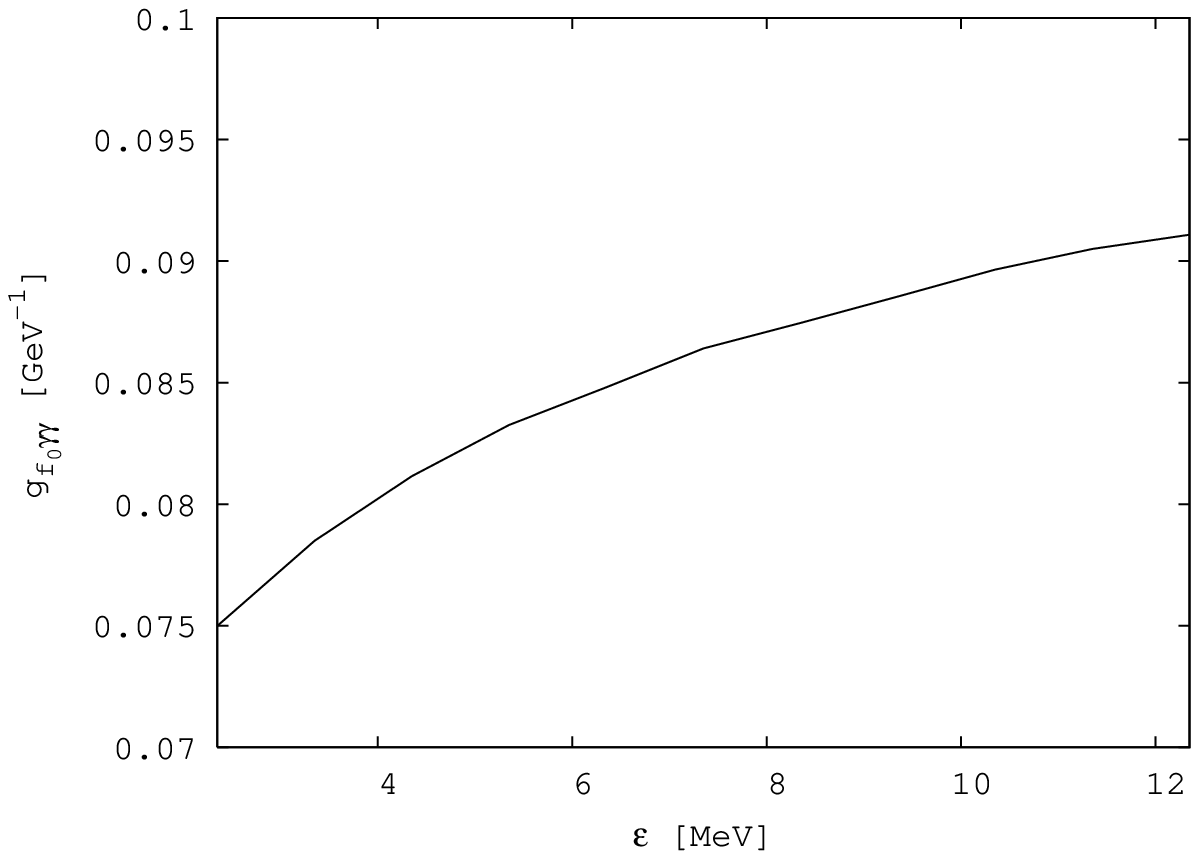,height=9cm}}
\caption{Coupling constant $g_{f_0 \gamma \gamma}$ in the local case \
in dependence on $\epsilon$.}

\vspace*{1.5cm} 

\centering{\
\epsfig{figure=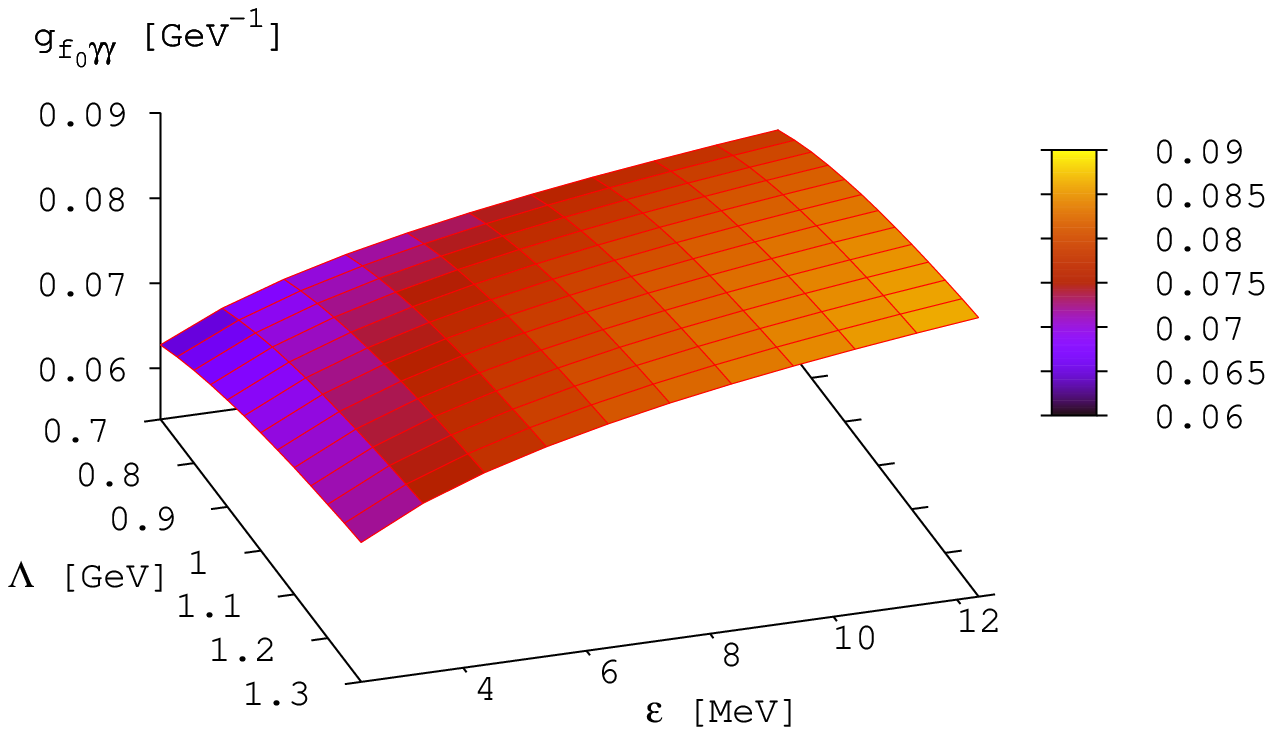,height=10cm}}
\caption{Coupling constant $g_{f_0 \gamma \gamma}$ 
in the nonlocal case in dependence on $\epsilon$ and $\Lambda$.}
\end{figure*}

\newpage
\begin{figure*}
\centering{\
\epsfig{figure=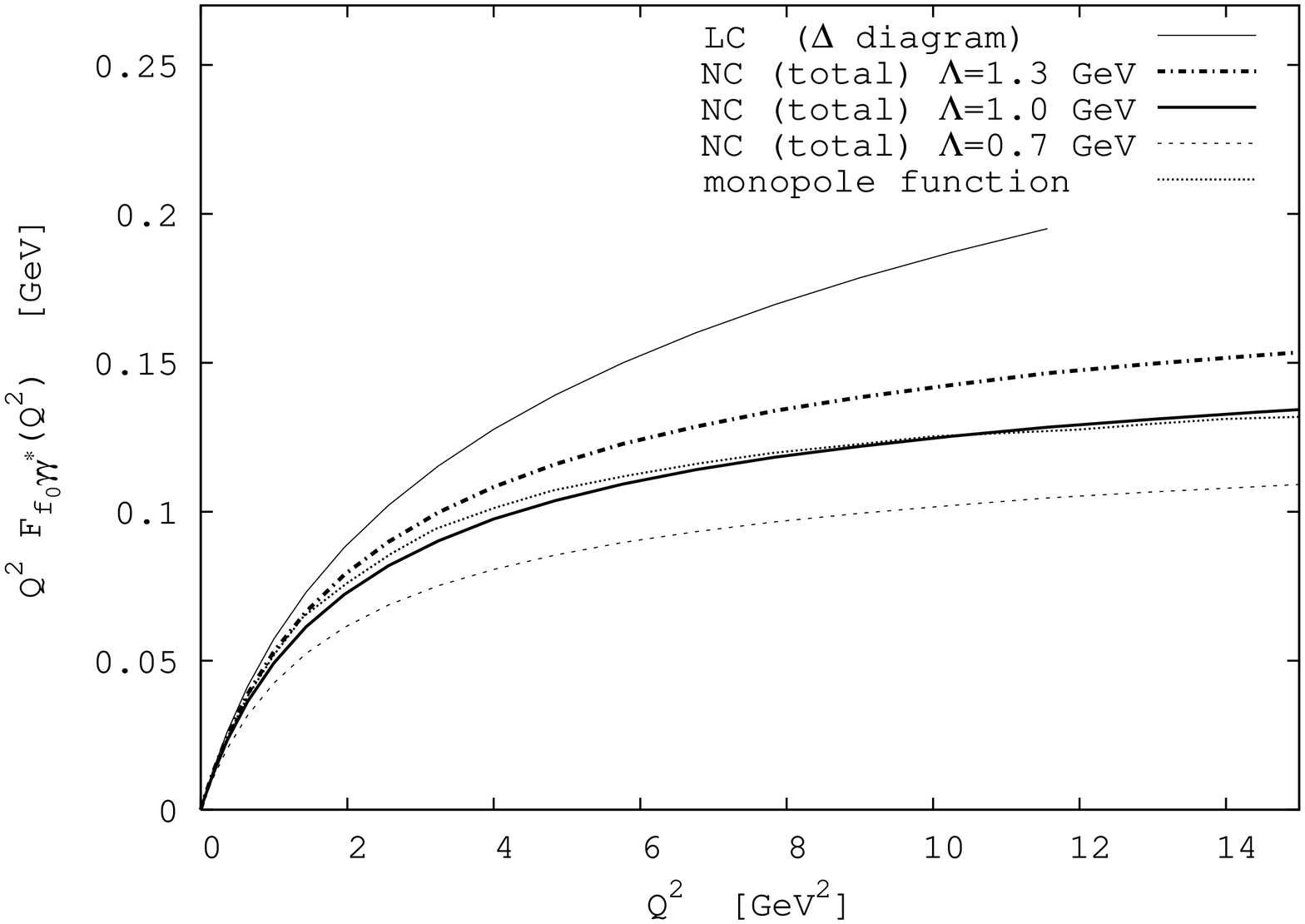,height=11cm,angle=-90}
\vspace*{.5cm} 
\caption{The form factor $Q^2F_{f_0\gamma\gamma^\ast}(Q^2)$ 
in dependence on $Q^2$ for the local case (LC). 
For the nonlocal case (NC) results are given for the 
triangle ($\Delta$) diagram and for all of Fig.~4 (total). 
The binding energy is set to $\epsilon$=7.4 MeV for all curves. 
For the monopole function see the text.}}
\vspace*{.5cm} 

\centering{\
\epsfig{figure=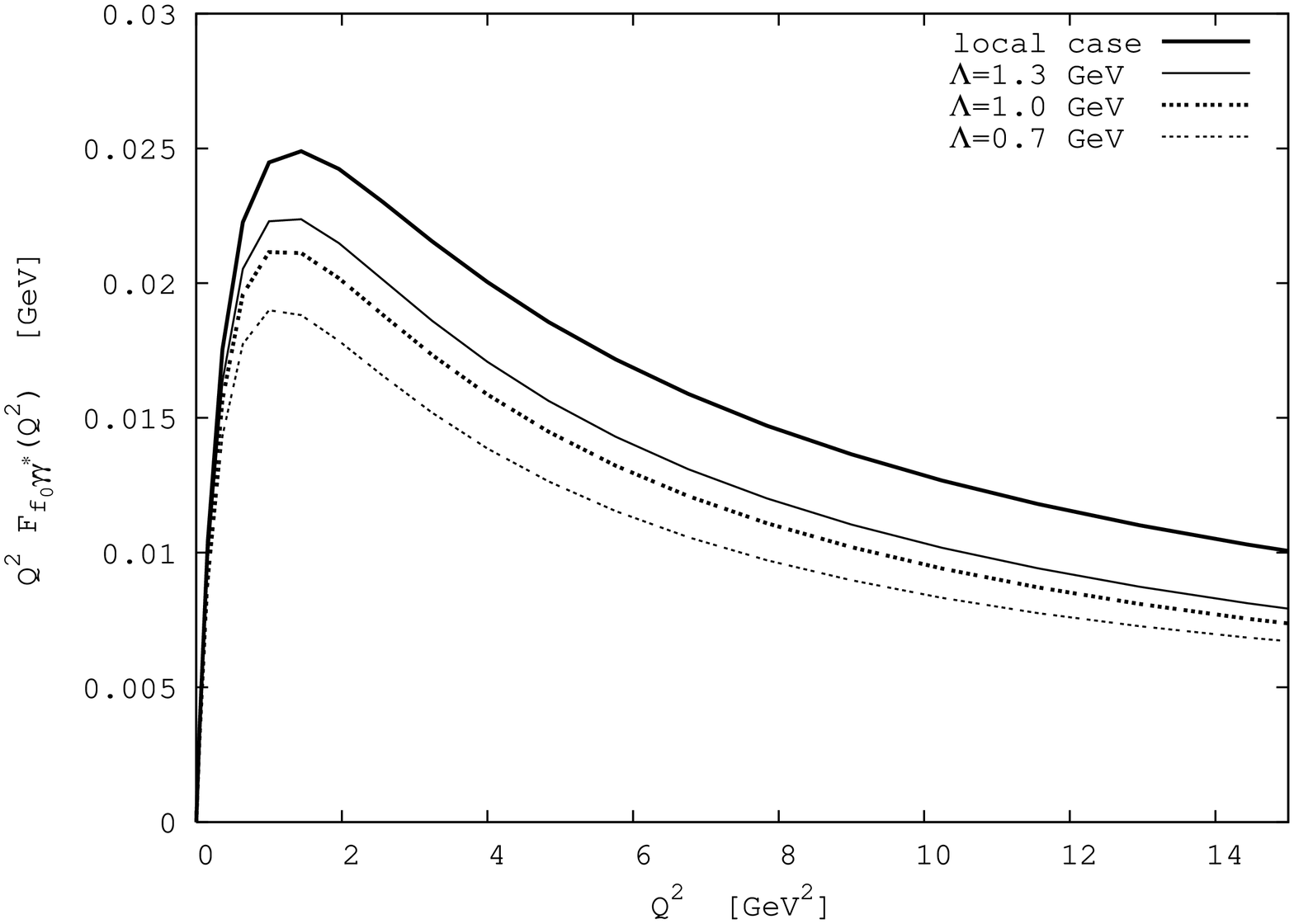,height=11cm,angle=-90}
\caption{The form factor $Q^2F_{f_0\gamma\gamma^\ast}(Q^2)$ 
in dependence on $Q^2$ with the additional monopole form factors $F_{K\bar K\gamma}(Q^2)$ at the $K\bar K\gamma$ vertices.}}
\label{fig:ff}
\end{figure*}
\newpage
 \begin{figure*}
\centering{\
\epsfig{figure=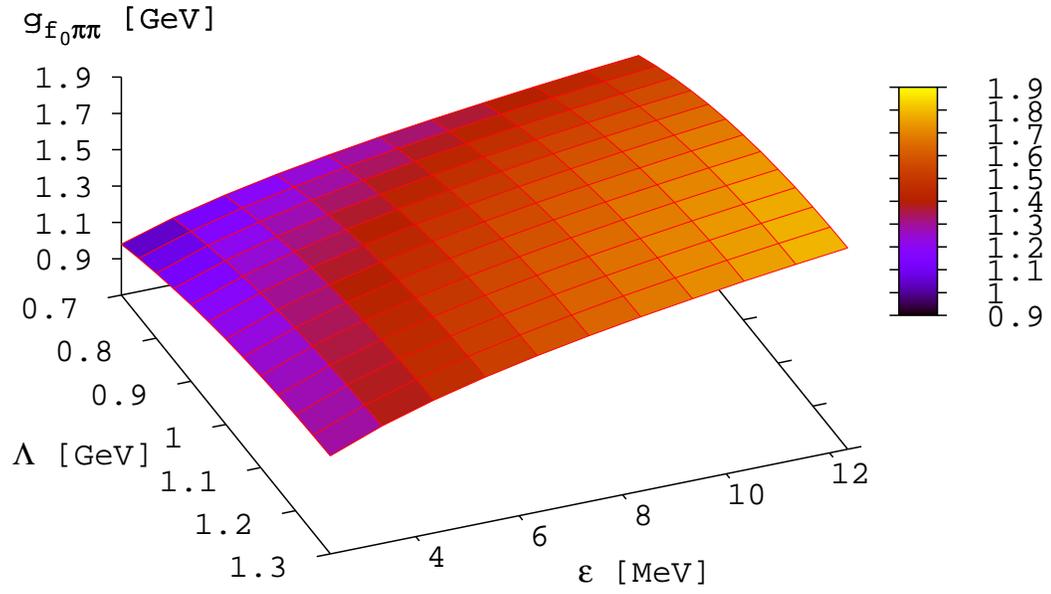,height=10cm}
\caption{Coupling constant $g_{f_0 \pi \pi}$ in the local case.}}
\label{fig:g}
\end{figure*}

\end{document}